\documentclass[aps,prl,preprintnumbers,showpacs,nofootinbib]{revtex4}
\usepackage{graphicx}% Include figure files
\usepackage{dcolumn}% Align table columns on decimal point
\usepackage{bm}% bold math
\begin{document}
\newcommand{\newc}{\newcommand}
\newcommand{\psla}{\not \! p}
\newc{\ra}{\rightarrow}
\newc{\lra}{\leftrightarrow}
\newc{\beq}{\begin{equation}}
\newc{\eeq}{\end{equation}}
\newc{\barr}{\begin{eqnarray}}
\newc{\earr}{\end{eqnarray}}
%%%%%%%%%%%%%%%%%%%%%%%%%%%%%%%%%%%%%%%%%%%
\newcommand{\Od}{{\cal O}}
\newcommand{\lsim}   {\mathrel{\mathop{\kern 0pt \rlap
  {\raise.2ex\hbox{$<$}}}
  \lower.9ex\hbox{\kern-.190em $\sim$}}}
\newcommand{\gsim}   {\mathrel{\mathop{\kern 0pt \rlap
  {\raise.2ex\hbox{$>$}}}
  \lower.9ex\hbox{\kern-.190em $\sim$}}}
  \def\rpm{R_p \hspace{-0.8em}/\;\:}
%\preprint{APS/123-QED}
%\date{\today}
\title {DIRECT DETECTION OF DARK MATTER RATES FOR VARIOUS WIMPS}
%\title {EXPLOITING  TPC DETECTORS AND THE COHERENT  NEUTRAL CURRENT INTERACTION
%FOR DETECTING  SUPERNOVA NEUTRINOS}
%
%
%\toctitle{ Neutrinos \protect\newline in a Spherical Box}
% allows explicit linebreak for the table of content
%
%
%\titlerunning{NEUTRINOS IN A SPHERICAL BOX}
% allows abbreviation of title, if the full title is too long
% to fit in the running head
%
%
\author{V.K. Oikonomou$^{1}$, J.D. Vergados$^{2}$\footnote{e-mail: vergados@cc.uoi.gr} and   Ch.C. Moustakidis$^{1}$
%\footnote{e-mail: moustaki@auth.gr}
}

\affiliation{ $^1$  Department of Theoretical Physics, Aristotle
University of Thessaloniki, 54124 Thessaloniki, Greece \\
$^2$ University of Ioannina, Ioannina, GR 45110, Greece.}
%\\E-mail:Vergados@cc.uoi.gr
%\maketitle              % typesets the title of the contribution
%\begin{frontmatter}
\vspace{0.5cm}
\vspace{0.5cm}
\begin{abstract}
{\bf The event rates for the direct detection of dark matter
candidates, originating from UED scenario, are evaluated for a
number of nuclear targets. Realistic form factors as well as spin
ME and response functions are employed. Due to LR+RL helicities
contribution, the proton amplitude is found to be dominant.
Various other non-susy dark matter candidates are examined at the
end.}
%The event rates for the direct detection of a dark matter
%candidate, which is a Kaluza-Klein gauge boson, are evaluated for
%a number of nuclear targets. Realistic form factors as well as
%spin ME and response functions are employed.
\end{abstract}
 %\end{document}
%\begin{keyword}
\pacs{95.35+d, 12.60.Jv.}
%%%%%%%%%%%%%%%%%%%%%%%%%%%%%%%%%%%%%%%%%%%%%%%%%%%%%%%%%%%%%%%%%%%%%
%\date{\today}
\maketitle
%\end{keyword}
%\end{frontmatter}
%%%%%%%%%%%%%%%%%%%%%%%%%%%%%%%%%%%%%%%%%%%%%%%%%%%%%%%%%%%%%%%%%%%%%
%%%%%%%%%%%%%%%%%%%%%%%%%%%%%%%%%%%%%%%%%%%%%%%%%%%%%%%%%%%%%%%%%%%%%
\section{Introduction.}
Models with compact extra dimensions offer rich and interesting
phenomenology
\cite{ST02a,ST02b,Antoniadis-a,Arkani-b,Dienes-c,Appelquist-d,Antoniadis-g,Antoniadis-h}.
In such models fields propagating in extra dimensions at low
energies appear as a tower of massive particles corresponding to a
given charge and spin. The massive states are nothing but modes of
the fields carrying quantized momentum in extra dimensions. This
means that the spacing of the towers is $\frac{1}{R}$, i.e. the
inverse of the characteristic size in extra dimensions. In this
scheme the ordinary particles are associated with the zero modes.
In brane world models only fields interacting gravitationally can
propagate in extra dimensions, i.e. the excitations are of the
Kaluza-Klein (K-K) type. Models with Universal Extra Dimensions
can have stable particles, { due to KK parity, originating from
higher dimensional Poincare invariance }\cite{Appelquist-d}. Under
this parity the even modes,  including the ordinary particles, are
even and the odd modes are odd. Thus the lightest odd mode
particle is cosmologically stable. For a recent review we refer
the reader to the literature \cite{SERVANT}. Like the neutralino,
it must be a neutral a weakly interacting particle. It can thus
serve as a viable dark matter candidate, which, together with dark
energy,
 seems to dominate in the Universe \cite{MAXIMA-1,BOOMERANG,DASI,COBE,SPERGEL,SDSS,WMAP06}.

The K-K WIMP's (Weekly Interacting Massive Particles) velocity
dependence can be assumed to be the same with that used in
neutralino calculations, since a particle's rotational velocity is
independent of  its mass.  The kinematics involved is, by and
large, similar to that of the neutralino, leading to cross
sections which are proportional $\mu_r$, the WIMP-nucleus reduced
mass. Furthermore the nuclear physics input, which depends on
$\mu_r\simeq Am_p$, is expected to  be the same. There are appear
two differences compared to the neutralino, though,  both related
to its larger mass.
\begin{itemize}
\item The density (number of particles per unit volume) of a WIMP
falls inversely proportional to its mass. Thus, since in K-K
theories the WIMP mass is much larger than that of the target, for
a given WIMP-nucleon gross section, the event rate is inversely
proportional to the WIMP mass. This means that the limits on the
nucleon cross section extracted from the  data must be rising with
the square root of the WIMP mass. This allows for nucleon
 cross sections larger than those extracted for the neutralino.
\item The average neutralino energy is now quite higher. In fact
for a Maxwell-Boltzmann (M-B) velocity distribution one finds that
$< T_{WIMP}> =\frac{3}{4}M_{WIMP} \upsilon^2_0$, with $\upsilon_0$
the characteristic velocity of the M-B distribution, which
coincides with the sun's rotational velocity, $\upsilon_0\simeq
2.2\times 10^5$km/s. Thus one finds
$$<
T_{WIMP}>\simeq 40 \left (\frac{m_{WIMP}}{100 \mbox{
GeV}}\right )\mbox{ keV}.$$
Furthermore, since the maximum allowed velocity is $\upsilon_{esc}=2.84 \upsilon_0$, we find that
$$T_{max}\simeq 120 \left (\frac{m_{WIMP}}{100 \mbox{
GeV}}\right )\mbox{ keV}.$$
 Thus for a K-K WIMP with mass $1$ TeV, the
average WIMP energy is $0.4$ MeV and the maximum energy is $1.2$ MeV. Thus in this case, due to the
high velocity tail of the velocity distribution it is reasonable
to expect { an energy transfer to the nucleus  in the MeV
region. So one need not attempt to detect such a heavy WIMP  the
hard way, i.e. by  measuring the energy of the recoiling nucleus,
as in the case of the neutralino. Many nuclear targets can now be
excited by the WIMP-nucleus interaction and the de-excitation
photons can be easily detected.}
 \end{itemize}
 \section{The Kaluza-Klein Boson as a dark matter candidate}
 \label{KK}
 { Assuming small boundary terms} we expect that the lightest exotic particle,
 which can serve as a dark matter candidate, is a gauge boson $B^{1}$
 having the same quantum numbers and couplings with the Standard Model gauge boson $B$, except that it has K-K parity
 $-1$. Thus its couplings involve another negative K-K parity particle. In this work we will assume that such  particles
 are the K-K quarks, partners of the ordinary quarks, but much heavier \cite{ST02a,ST02b}.
\subsection{Intermediate K-K quarks}
\begin{figure}[!ht]
\begin{center}
\includegraphics[scale=1.0]{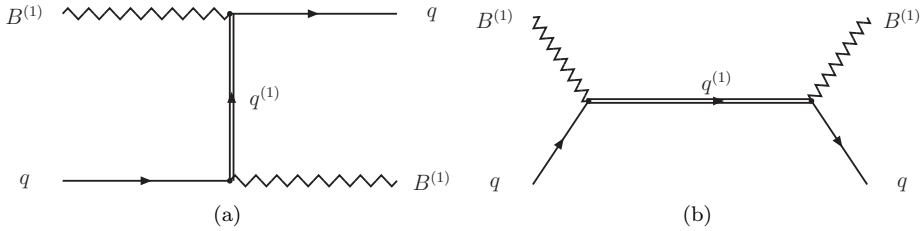}
%\hspace*{0.0cm}\tiny{$(T_A)_{th} \rightarrow$MeV }
 \caption{Two diagrams leading to the interaction of K-K gauge boson $B^{1}$ with quarks at tree
level mediated by K-K quarks}
 \label{fig:kkq}
 \end{center}
  \end{figure}

In this case a generic Feynman diagrams are
 shown in Fig.~\ref{fig:kkq}.
 The corresponding amplitude involving left-handed quarks is given by:
 \barr
{\cal M}_1^{L}(q)&=& -i(g_1 Y_{L})^2 \left[
\overline{q_L}(x) \gamma^{\nu}
\frac{(\psla_q-\psla_{B^{\prime}}+m_{q^{(1)}})}
{(p_q-p_{B^{\prime}})^2-m^2_{q^{(1)}}}
\gamma^{\mu}  q_L(x) \right]
\epsilon^*_{\mu}(p^\prime_B) \epsilon_{\nu}(p_{B}), \hspace*{0.5cm}
\label{eq:kkq1}\\
{\cal M}_2^{L}(q)&=& -i(g_1 Y_{L})^2 \left[ \overline{q_L}(x)
\gamma^{\mu} \frac{(\psla_q+\psla_{B}+m_{q^{(1)}})}
{(p_q+p_B)^2-m^2_{q^{(1)}}} \gamma^{\nu}  q_L(x) \right]
\epsilon^*_{\mu}(p_{B^{\prime}}) \epsilon_{\nu}(p_B)
,\hspace*{0.5cm} \label{eq:kkq2} \earr where $g_1=g
\tan{\theta_W}=\sqrt{4 \sqrt{2}G_F}m_W \tan{\theta_W}$, $Y_L=1/3$
and $\epsilon_{\mu}(p^\prime_B),\epsilon_{\nu}(p_{B})$ are the
helicities of the K-K bosons. For the right handed quarks of the
upper type we have,
 \barr
{\cal M}_1^{R}(u)&=& -i(g_1 4/3)^2 \left[
\overline{u_R}(x) \gamma^{\nu}
\frac{(\psla_q-\psla_{B^{\prime}})}
{(p_q-p_{B^{\prime}})^2-m^2_{q^{(1)}}}
\gamma^{\mu}  u_R(x) \right]
\epsilon^*_{\mu}(p^\prime_B) \epsilon_{\nu}(p_{B}), \hspace*{0.5cm} \\
{\cal M}_2^{R}(u)&=& -i(g_1 4/3)^2 \left[ \overline{u_R}(x)
\gamma^{\mu} \frac{(\psla_q+\psla_{B})}
{(p_q+p_B)^2-m^2_{q^{(1)}}} \gamma^{\nu}  u_R(x) \right]
\epsilon^*_{\mu}(p_{B^{\prime}}) \epsilon_{\nu}(p_B)
.\hspace*{0.5cm} \earr For the right handed down quarks  we have,
 \barr
{\cal M}_1^{R}(d)&=& -i(g_1(- 2/3))^2 \left[
\overline{d_R}(x) \gamma^{\nu}
\frac{(\psla_q-\psla_{B^{\prime}})}
{(p_q-p_{B^{\prime}})^2-m^2_{q^{(1)}}}
\gamma^{\mu}  d_R(x) \right]
\epsilon^*_{\mu}(p^\prime_B) \epsilon_{\nu}(p_{B}), \hspace*{0.5cm} \\
{\cal M}_2^{R}(d)&=& -i(g_1 (-2/3))^2 \left[ \overline{u_R}(x)
\gamma^{\mu} \frac{(\psla_q+\psla_{B})}
{(p_q+p_B)^2-m^2_{q^{(1)}}} \gamma^{\nu}  d_R(x) \right]
\epsilon^*_{\mu}(p_{B^{\prime}}) \epsilon_{\nu}(p_B)
.\hspace*{0.5cm} \earr We also have R-L interference terms, which
seem to have been missed in the previous calculations \cite{ST02a,ST02b}.
One finds:
 \barr
{\cal M}_1^{LR}(u)&=& -i(g_1)^2 4/9 \left[
\overline{u_L}(x) \gamma^{\nu}
\frac{m_{q^{(1)}}}
{(p_q-p_{B^{\prime}})^2-m^2_{q^{(1)}}}
\gamma^{\mu}  u_R(x) +H.C.\right]
\epsilon^*_{\mu}(p^\prime_B) \epsilon_{\nu}(p_{B}), \hspace*{0.5cm} \\
{\cal M}_2^{LR}(u)&=& -i(g_1)^2 4/9 \left[
\overline{u_L}(x) \gamma^{\mu}
\frac{m_{q^{(1)}}}
{(p_q+p_B)^2-m^2_{q^{(1)}}}
\gamma^{\nu}  u_R(x) +H.C. \right]
\epsilon^*_{\mu}(p_{B^{\prime}}) \epsilon_{\nu}(p_B) ,\hspace*{0.5cm}
\earr
 \barr
{\cal M}_1^{LR}(d)&=& -i(g_1)^2 (-2/9) \left[
\overline{d_L}(x) \gamma^{\nu}
\frac{m_{q^{(1)}}}
{(p_q-p_{B^{\prime}})^2-m^2_{q^{(1)}}}
\gamma^{\mu}  d_R(x) +H.C.\right]
\epsilon^*_{\mu}(p^\prime_B) \epsilon_{\nu}(p_{B}), \hspace*{0.5cm} \\
{\cal M}_2^{LR}(d)&=& -i(g_1)^2 (-2/9) \left[ \overline{d_L}(x)
\gamma^{\mu} \frac{m_{q^{(1)}}} {(p_q+p_B)^2-m^2_{q^{(1)}}}
\gamma^{\nu}  d_R(x) +H.C. \right]
\epsilon^*_{\mu}(p_{B^{\prime}}) \epsilon_{\nu}(p_B)
.\hspace*{0.5cm} \earr
Since both the K-K bosons and quarks are
very massive and the momenta of the external particles are quite
small one can employ the non relativistic limit. Thus the
helicities of the K-K bosons are space like. Thus to leading order
the amplitude corresponding to Eqs (\ref {eq:kkq1}) and (\ref
{eq:kkq2}) can be written: \barr {\cal M}_q^{L}&=& -ig_1^2 (1/9)
\overline{q}(x)\frac{1}{2}\left [(-\mbox{\boldmath
$\epsilon^{*'}$}.\mbox{\boldmath $\epsilon$}) \gamma^0
-i(\mbox{\boldmath $\epsilon^{*'}$}\times \mbox{\boldmath
$\epsilon$}).\mbox{\boldmath $\gamma$} \gamma_5 \right ]q(x)\times
\nonumber \\
& &\left\{  \frac{(E_q+m_{B^{(1)}}) }
{(m_{B^{(1)}}+E_q)^2-m^2_{q^{(1)}}} +\frac{(E_q-m_{B^{(1)}})}
{(m_{B^{(1)}}-E_q)^2-m^2_{q^{(1)}}} \right\} . \earr
The first
term in the square bracket is spin independent and it can lead to
coherence. The second term depends on the spin and it can not lead
to coherence. It may be important only in the case of odd nuclear
targets. The curly bracket in the last equation can be brought
into the form:
$$\frac{E_q}{(m_{B^{(1)}})^2} f_1(\Delta )~, ~f_1(\Delta )=\frac{1+\Delta +\Delta ^2 /2}{\Delta ^2(1+\Delta /2)^2}~,
~\Delta =\frac{m_{q^{(1)}}}{m_{B^{(1)}}}-1.$$ We see that the
amplitude is very sensitive to the parameter $\Delta $ ("resonance
effect"). For values of $\Delta$ not very close to zero the event
rate is perhaps unobserved since the mass of the K-K boson is
expected to be large.
 In the case of the right handed interaction we obtain an analogous expression:
\beq {\cal M}_u^{R}= -ig_1^2 (16/9)
\overline{u}(x)\frac{1}{2}\left [(-\mbox{\boldmath
$\epsilon^{*'}$}.\mbox{\boldmath $\epsilon$}) \gamma^0
+i(\mbox{\boldmath $\epsilon^{*'}$}\times \mbox{\boldmath
$\epsilon$}).\mbox{\boldmath $\gamma$} \gamma_5 \right
]u(x)\frac{E_q}{(m_{B^{(1)}})^2} f_1(\Delta ) ,\eeq \beq {\cal
M}_d^{R}= -ig_1^2 (4/9) \overline{d}(x)\frac{1}{2}\left
[(-\mbox{\boldmath $\epsilon^{*'}$}.\mbox{\boldmath $\epsilon$})
\gamma^0 +i(\mbox{\boldmath $\epsilon^{*'}$}\times \mbox{\boldmath
$\epsilon$}).\mbox{\boldmath $\gamma$} \gamma_5 \right
]d(x)\frac{E_q}{(m_{B^{(1)}})^2} f_1(\Delta ). \eeq In the case of
the R-L interference term there is no $\gamma_5$ term. Furthermore
the amplitude to leading order is now independent of the energy of
the quark. We thus find: \beq {\cal M}_u^{Rl}= -ig_1^2 (4/9)
\overline{u}(x) \left [(-\mbox{\boldmath
$\epsilon^{*'}$}.\mbox{\boldmath $\epsilon$})  \right
]u(x)\frac{1}{(m_{B^{(1)}})} f_2(\Delta ), \eeq \beq {\cal
M}_d^{Rl}= -ig_1^2 (-2/9) \overline{d}(x) \left [(-\mbox{\boldmath
$\epsilon^{*'}$}.\mbox{\boldmath $\epsilon$})  \right
]d(x)\frac{1}{(m_{B^{(1)}})} f_2(\Delta ), \eeq with
$$~f_2 (\Delta )=\frac{1+\Delta }{\Delta (1+\Delta /2)}.$$
The results in this case are less sensitive to $\Delta $.

\begin{figure}[!ht]
 \begin{center}
\includegraphics[scale=1.0]{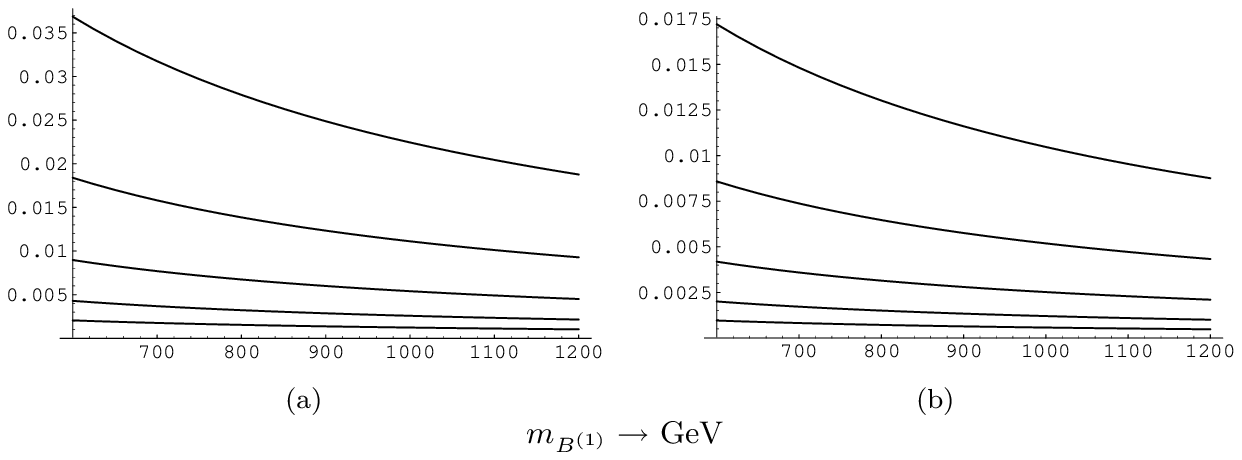}
 \hspace{0.0cm} $m_{B^{(1)}}\rightarrow$ GeV
%\hspace*{0.0cm}\tiny{$(T_A)_{th} \rightarrow$MeV }
 \caption{On the left we show the ratio $(LL+RR)/(LL+RR+LR+RL)$ of the various chirality amplitudes
in the case of the proton.  On the right we we show the
the ratio of the amplitude of the neutron divided by that of the proton. Both are exhibited as functions
of $m_{B^{(1)}}$ in GeV for the values of $\Delta=0.05,0.10,0.20$ and $0.40$ (from top to bottom). We see that
due to the LR+RL contribution the amplitude associated with the proton is dominant.
}
 \label{fig:ratio}
  \end{center}
  \end{figure}

The next step involves going from the quark to the nucleon level.
The only question concerns the quark energy.
 It seems to us that the best procedure is to replace the quark energy with their constituent mass $\simeq 1/3m_p$, as
opposed to adopting \cite{ST02a,ST02b} a procedure related to the current mass encountered in the neutralino
case \cite{Chen,Dree00},\cite{JDV06}. In the latter case the amplitude at the quark level is proportional to the quark mass
(see the intermediate Higgs exchange below). Thus in this case, unlike the neutralino dark matter,
 the process is dominated
by the quarks u and d. So the obtained  results do not critically
depend on the quark content of the nucleon. We thus find: \beq
{\cal M}_{coh}= i 4 \sqrt{2} G_F m_W \tan^2{\theta_W
}(\mbox{\boldmath $\epsilon^{*'}$}.\mbox{\boldmath
$\epsilon$})N\left [ \left ( \frac{11}{18}+\frac{2}{3}\tau_3
\right ) \frac{1}{3} \frac{m_p m_W}{(m_{B^{(1)}})^2} f_1(\Delta
)+(\frac{1}{3}+\frac{1}{3}\tau_3 ) \frac{m_W}{ m_{B^{(1)}}}
f_2(\Delta )  \right ] N .\eeq In Fig. \ref{fig:ratio} we present
the ratio of the amplitude arising from LL and RR contributions
alone divided by the total (LL+RR+LR+RL) in the case of the
proton. We see that the second term dominates even slightly away
from the resonance condition.

In the case of the spin contribution we find at the quark level
that: \beq {\cal M}_{spin}= -i 4 \sqrt{2} G_F m_W \tan^2{\theta_W
}\frac{1}{3} \frac{m_p m_W}{(m_{B^{(1)}})^2} f_1(\Delta )\bar{q}
i(\mbox{\boldmath $\epsilon^{*'}$}\times \mbox{\boldmath
$\epsilon$}).\left [  \frac{17}{18}\bar{u}\mbox{\boldmath
$\gamma$} \gamma_5 u + \frac{5}{18}\bar{d}\mbox{\boldmath
$\gamma$} \gamma_5 d + \frac{5}{18}\bar{s}\mbox{\boldmath
$\gamma$} \gamma_5 s  \right ] .\eeq In going to the nucleon level
we get the isoscalar part \cite{JDV06}
$$g_0=\frac{17}{18}\Delta u+\frac{5}{18} \Delta d+\frac{5}{18} \Delta s,$$
while the isovector part is:
$$g_1=\frac{17}{18}\Delta u-\frac{5}{18} \Delta d .$$
The quantities $\Delta_q$ are given by \cite{ST02a,ST02b,JDV06}
$$\Delta u=0.78\pm 0.02~,~\Delta d=-0.48\pm 0.02~,~\Delta s=-0.15\pm 0.02 .$$
We thus find,
$$g_0=0.26~,~g_1=0.41 .$$
In the proton neutron representation we obtain:
$$a_p=0.67~,~a_n=-0.15 .$$
The picture is different for the neutralino case in which \cite{JELLIS},
$$g_0=\Delta u+ \Delta d+ \Delta s=0.15~,~g_1=\Delta u- \Delta d=1.26~,~ a_p=1.41~,~a_n=-1.11.$$
Thus at the nucleon level we get \beq {\cal M}_{spin}= -i 4
\sqrt{2} G_F m_W \tan^2{\theta_W }\frac{1}{3} \frac{m_p
m_W}{(m_{B^{(1)}})^2} f_1(\Delta )\bar{q} i(\mbox{\boldmath
$\epsilon^{*'}$}\times \mbox{\boldmath $\epsilon$}).\left [
N\mbox{\boldmath $\sigma$} (g_0+g_1 \tau_3) N   \right ] .\eeq
%%%%%%%%%%%%%%%%%%%%%%%%%%%%%%%%%%%%%%%%%%%
\subsection{Intermediate Higgs Scalars}
\begin{figure}[!ht]
\begin{center}
\includegraphics[scale=1.0]{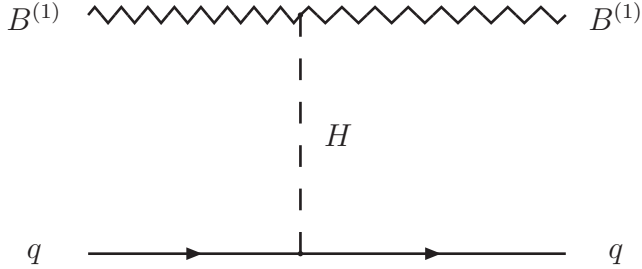}
%\hspace*{0.0cm}\tiny{$(T_A)_{th} \rightarrow$MeV }
\caption{The interaction of K-K gauge boson $B^{(1)}$ with quarks
at tree level mediated by Higgs scalars}
 \label{fig:kkh}
 \end{center}
  \end{figure}

%%%%%%%%%%%%%%%%%%%%%%%%%%%%%%%%%%%%%%%%%%%%
The corresponding Feynman diagram is shown in Fig. \ref{fig:kkh}
  The vertex involving the interaction of the Higgs particle with
  the K-K boson is given by
  \beq
  {\cal L}_{BBh}= g_1^2 \frac{1}{4}\epsilon^*_{\nu}(p^\prime_B)
  \epsilon_{\nu}(p_{B}) H H \Rightarrow  g_1^2 \frac{1}{4}\epsilon^*_{\nu}(p^\prime_B)
  \epsilon_{\nu}(p_{B}) h \prec H\succ
 . \eeq
  The coupling of the Higgs scalar to the quark is given by:
  \beq
  {\cal L}_{qqh}= \frac{m_q}{\prec H\succ} h
  .\eeq
  We thus obtain
\beq {\cal M}_q(h)= -ig_1^2 \frac{1}{4} \left [-\mbox{\boldmath
$\epsilon^{*'}$}.\mbox{\boldmath
$\epsilon$}~~\overline{q}(x)\frac{m_q}{m^2_h} q(x) \right ]
 .\eeq
 In going from the quark to the nucleon level we follow a procedure analogous to that of the of the
neutralino \cite{Dree00,Chen,JDV06},
  i.e.
  $$\prec  N |m_q~ q \bar{q}|N  \succ  \Rightarrow f_q m_p ,$$
  we thus get
  \beq
 {\cal M}_N(h)= -i~4 \sqrt{2}G_F m^2_W \tan^2{\theta_W}~\left [\frac{1}{4}\frac{m_p}{m^2_h} \left (-\mbox{\boldmath
$\epsilon^{*'}$}.\mbox{\boldmath
$\epsilon$} \right ) \prec N|N\succ  \sum_q f_q\right ]
  .\eeq
  In this case the proton and the neutron cross sections are about equal.
%%%%%%%%%%%%%%%%%%%%%%%%%%%%%%%%%%%%%%%%%%%%%%%%%%%%%%%%%%%%%%%%%%%%%%%%%%%%%%%%%%
\section{K-K neutrinos as dark matter candidates}
%%%%%%%%%%%%%%%%%%%%%%%%%%%%%%%%%%%%%%%%%%%%%%%%%%%%%%%%%%%%%%%%%%%%%%
The other possibility is the dark matter candidate to be a heavy
K-K neutrino. We distinguish the following cases
%%%%%%%%%%%%%%%%%%%%%%%%%%%%%%%%%%%%%%%%%%%%%
\subsection{Process mediated by Z-exchange}
\begin{figure}[!ht]
 \begin{center}
\includegraphics[scale=1.0]{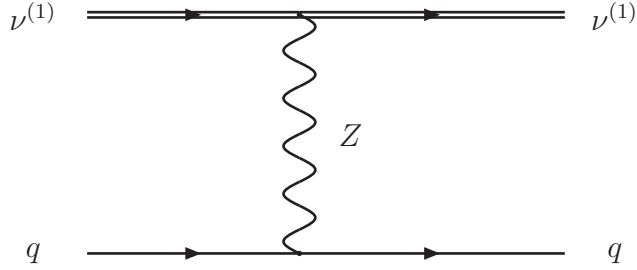}
%\hspace*{0.0cm}\tiny{$(T_A)_{th} \rightarrow$MeV }
 \caption{The interaction of K-K neutrino $\nu^{(1)}$ with quarks at tree
level mediated by Z-exchange}
 \label{fig:kknu}
 \end{center}
  \end{figure}

%%%%%%%%%%%%%%%%%%%%%%%%%%%%%%%%%%%%%%%%%%%%%%
 The Feynman diagram associated
 with this process is shown in Fig. \ref{fig:kknu}.
  The $qqZ$ vertex is given by,
  \beq
  f_{\lambda}=-\frac{1}{2}\frac{g}{2\cos{\theta_W}}J_{\lambda}(qqZ)
  ,\eeq
  \beq
  J_{\lambda}(qqZ)=\bar{q}\gamma_{\lambda} \left [-2 \sin^2{\theta_W} (\frac{1}{3}+\tau_3)+(1-\gamma_5)\tau_3    \right]q
  ,\eeq
    \beq
  J_{\lambda}(NNZ)= \bar{N}\gamma_{\lambda} \left [-2\sin^2{\theta_W} (1+\tau_3)+(1-g_A \gamma_5)\tau_3    \right]N
  .\eeq
 Thus in the case for the proton we encounter the combination,
 $$(-g_A \gamma_5+1-4 \sin^2{\theta_W})\simeq -g_A\gamma_5$$
 while in the case of the neutron:
 $$-1+g_A\gamma_5,$$
 which exhibit the well known fact the in the case of the neutral current interaction we can have coherence over the
 neutrons, but no coherence over the protons. Thus,
 \beq
J_{\lambda}(NNZ)\simeq \bar{N}\gamma_{\lambda} \left
[-g_A\gamma_5\tau_3-\frac{1}{2}(1-\tau_3 \right ]N . \eeq
 In the case of the neutrino vertex we write:
\beq f_{\lambda}(\nu^{(1)})=-\frac{1}{2}\frac{g}{2\cos{\theta_W}}
J_{\lambda}(\nu^{(1)}) .\eeq Regarding the neutrino current we now
have two possibilities:
 \begin{itemize}
 \item the K-K neutrino is a Majorana particle. In this case the neutrino current is:
 \beq
 J_{\lambda}(\nu^{(1)})= \bar{\nu}^{(1)}\gamma _{\lambda }\gamma_5\nu^{(1)}
 .\eeq
 The Majorana neutrino has no electromagnetic properties (no neutral vector current interaction).
 \item The K-K neutrino is a Dirac particle. In the presence of left handed interaction only we have:
  \beq
 J_{\lambda}(\nu^{(1)})= \bar{\nu}^{(1)}\gamma _{\lambda }(1-\gamma_5)\nu^{(1)}
 .\eeq
 \end{itemize}
 The amplitude associated with the diagram of Fig. \ref{fig:kknu} becomes:
 \beq
 {\cal M}_{\nu^{(1)}}=\frac{1}{4}\frac{g^2}{4 \cos^2{\theta_W}}\frac{1}{-m^2_Z}J^{\lambda}(\nu^{(1)}) J_{\lambda}(NNZ)=
-\frac{1}{2 \sqrt{2}}G_FJ^{\lambda}(\nu^{(1)}) J_{\lambda}(NNZ)
 .\eeq
 In other words in the case of the Majorana neutrino we get:
 \beq
 {\cal M}_{\nu^{(1)}}=\frac{1}{2 \sqrt{2}}G_F \bar{N}\gamma_{\lambda} g_A\gamma_5\tau_3 N\bar{\nu}^{(1)}\gamma _{\lambda }\gamma_5\nu^{(1)}
 ,\eeq
 while in the case of a Dirac neutrino the coherent contribution dominates, i.e.:
  \beq
 {\cal M}_{\nu^{(1)}}=\frac{1}{2 \sqrt{2}}G_F \bar{N}\gamma_{\lambda} \frac{1-\tau_3 }{2}N\bar{\nu}^{(1)}\gamma _{\lambda }
   (1-\gamma_5)\nu^{(1)}
 .\eeq
% \subsubsection{Process mediated by the Kaluza -Klein gauge boson $Z^{(1)}$}
% In this one may have production of the standard light neutrino (see Fig. \ref{fig:nu1nu} ). The amplitude is given
% as above.
%     \begin{figure}[!ht]
% \begin{center}
%\includegraphics[scale=0.8]{kknu1nu.eps}
%\hspace*{0.0cm}\tiny{$(T_A)_{th} \rightarrow$MeV }
% \caption{The interaction of  $\nu^{(1)}$-$\nu$ (K-K neutrino-ordinary neutrino) and quarks  mediated by the K-K $Z^{(1)}$}
% \label{fig:nu1nu}
% \end{center}
%  \end{figure}
%%%%%%%%%%%%%%%%%%%%%%%%%%%%%%%%%%%%%%%%%%%%%%%%%%%%%%%%%%%%%%%%%
\subsection{Process mediated by right handed currents  via $Z'$-boson exchange}
%%%%%%%%%%%%%%%%%%%%%%%%%%%%%%%%%%%%%%%%%%%%%%%%%%%%%%%%%%%%%%%
 The process is similar to that exhibited by Fig. \ref{fig:kknu}, except that instead of Z we encounter $Z'$, which is much
heavier. We will assume that the couplings of the $Z'$ are similar to those of $Z$.
Then the above results apply except that now the amplitudes are retarded by the multiplicative factor $\kappa={m^2_{Z}}/{m^2_{Z'}}$
\subsection{Process mediated by Higgs exchange}
 In this case in Fig.~\ref{fig:kknu},  Z is replaced by the Higgs particle. In this case the amplitude
at the quark level becomes:
  \beq
 {\cal M}_{\nu^{(1)}}(h)=
-2 \sqrt{2}G_F \frac{m_q m_{\nu^{(1)}}}{m_h^2}
\bar{\nu}^{(1)}~\nu^{(1)} \bar{q}~q . \eeq
 Proceeding as above we find that the amplitude at the nucleon level is:
  \beq
 {\cal M}_{\nu^{(1)}}(h)=
-2 \sqrt{2}G_F \frac{m_p m_{\nu^{(1)}}}{m_h^2}
\bar{\nu}^{(1)}~\nu^{(1)} \prec N|N \succ \sum_q f_q . \eeq
%%%%%%%%%%%%%%%%%%%%%%%%%%%%%%%%%%%%%%%%%%%%%%%%%%%%%%%%%%
\section{Nucleon cross sections}
%%%%%%%%%%%%%%%%%%%%%%%%%%%%%%%%%%%%%%%%%%%%%%%%%%%%%%%%%%%%%
In evaluating the nucleon cross section one proceeds as in the
case of the neutralino. The momentum transfer to the
 nucleon is $q=2 \mu_r \upsilon \xi$, where $\mu_r=\mbox{ reduced mass }\simeq m_p$, $\upsilon$ is the dark matter
 candidate velocity and $\xi$ is the cosine of the angle between the initial dark matter particle and the
outgoing nucleus.
%%%%%%%%%%%%%%%%%%%%%%%%%%%%%%%%%%%%%%%%%%%
\subsection{The K-K boson case}
%%%%%%%%%%%%%%%%%%%%%%%%%%%%%%%%%%%%%%%%%%%
To obtain the nucleon cross section, one must sum over the final
spin and boson polarizations and average over the initial ones.
One finds,
 \beq
 \sigma_N(coh)=\frac{1}{4 \pi}\frac{m_p^2}{(m_{B^{(1)}})^2} \frac{1}{2} \frac{1}{3}
 \sum_{pol,m_s}|{\cal M}_{coh}+{\cal M}_N(h)|^2
 ,\eeq
  \beq
 \sigma_N(spin)=\frac{1}{4 \pi}\frac{m_p^2}{(m_{B^{(1)}})^2} \frac{1}{2} \frac{1}{3}
 \sum_{pol,m_s}|{\cal M}_{spin}|^2
 ,\eeq
 where,
 $$\frac{1}{2} \frac{1}{3}
 \sum_{pol,m_s}=1~,~\frac{1}{2} \frac{1}{3}
 \sum_{pol,m_s}=6,$$
 for the spin independent and spin dependent parts respectively.

\begin{figure}[!ht]
 \begin{center}
\includegraphics[scale=1.0]{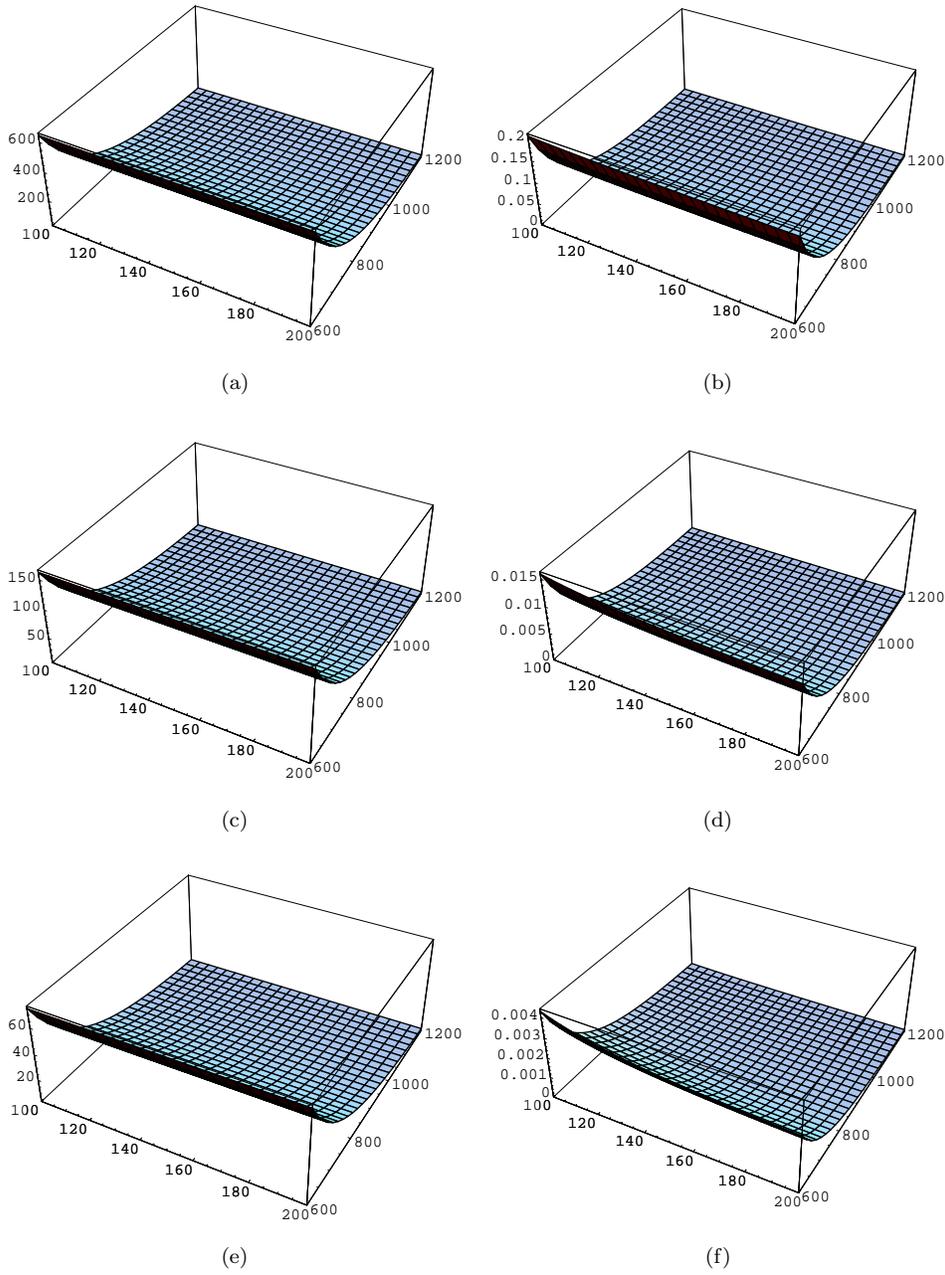}
%\hspace*{0.0cm}\tiny{$(T_A)_{th} \rightarrow$Me\hspace{-2.0cm} $m_{\chi}\rightarrow$ GeVV }
 \caption{The coherent proton cross section on the left and that for the neutron on the right in units of
$10^{-6}$pb, as a function of the gauge boson mass in the range of $600-1200$ GeV and the Higgs mass in
the range of $100-200$ GeV. From top to bottom $\Delta =0.05,0.10$ and $0.15$ respectively}
 \label{fig3d:coh}
  \end{center}
  \end{figure}

\begin{figure}[!ht]
 \begin{center}
 \includegraphics[scale=1.0]{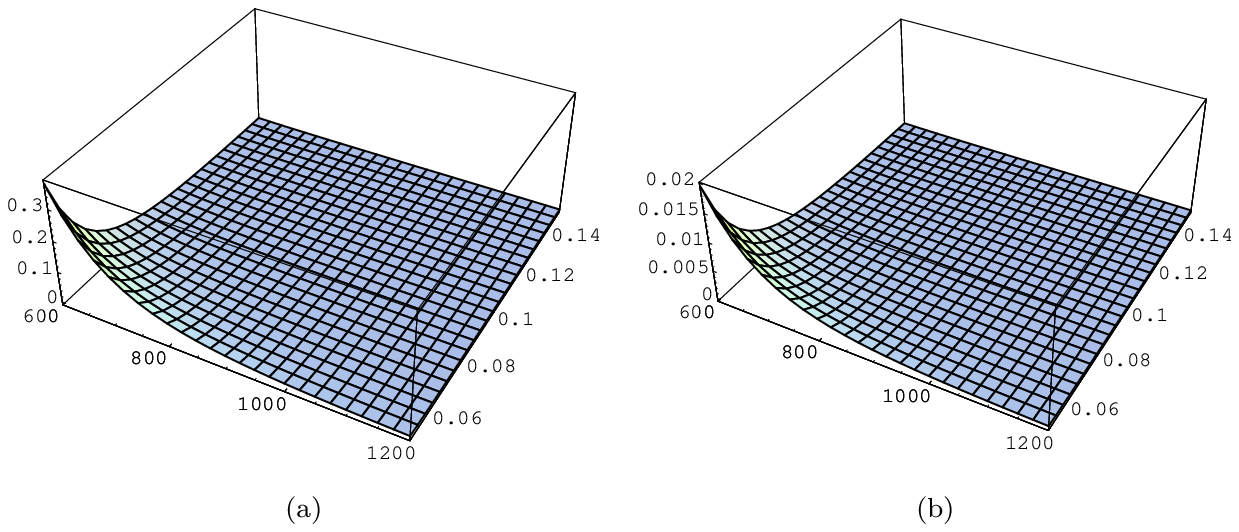}
 \caption{The spin proton cross section on the left and that for the neutron on the right in units of
$10^{-6}$pb, as a function of the gauge boson mass in the range of $600-1200$ GeV and $\Delta$ in the range
$0.05-0.15$.}
 \label{fig3d:spin}
 \end{center}
  \end{figure}

\begin{figure}[!ht]
 \begin{center}
 \includegraphics[scale=0.8]{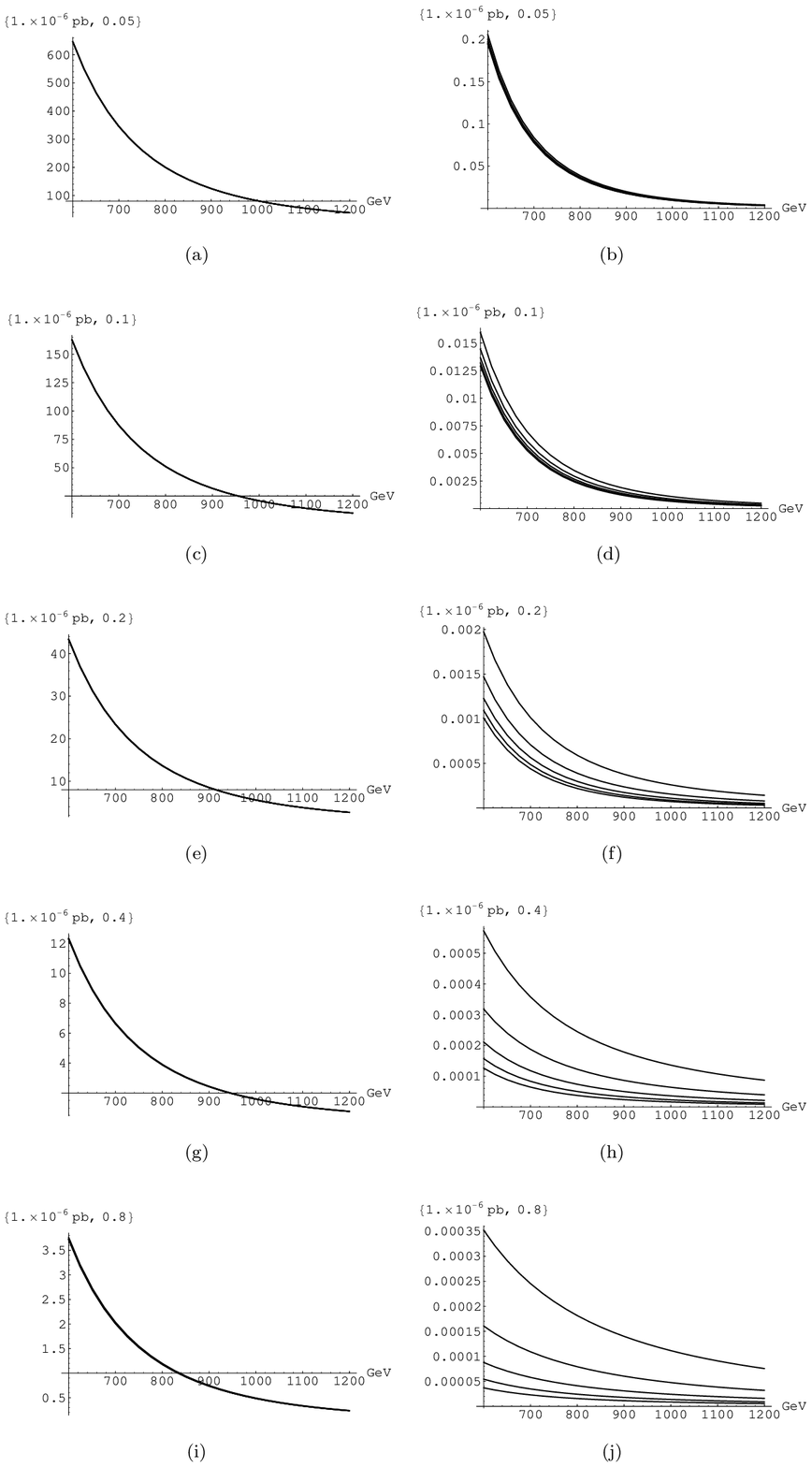}
%\hspace*{0.0cm}\tiny{$(T_A)_{th} \right
%\hspace*{0.0cm}\tiny{$(T_A)_{th} \rightarrow$MeV }
 \caption{The spin independent proton cross section on the left and that for the neutron on the right in units of
$10^{-6}$pb, as a function of the gauge boson mass in the range of $600-1200$ GeV.
>From top to bottom $\Delta =0.05,0.10,0.20,0.40$ and $0.80$. On each plot
we show results for $m_h=100,125,150,175$ and $200$ GeV with mass increasing downwards. Note that in the case of the proton,
due to the RL+LR dominance, the effect of Higgs contribution is not visible.}
 \label{fig2d:coh}
 \end{center}
  \end{figure}

\begin{figure}[!ht]
 \begin{center}
\includegraphics[scale=1.0]{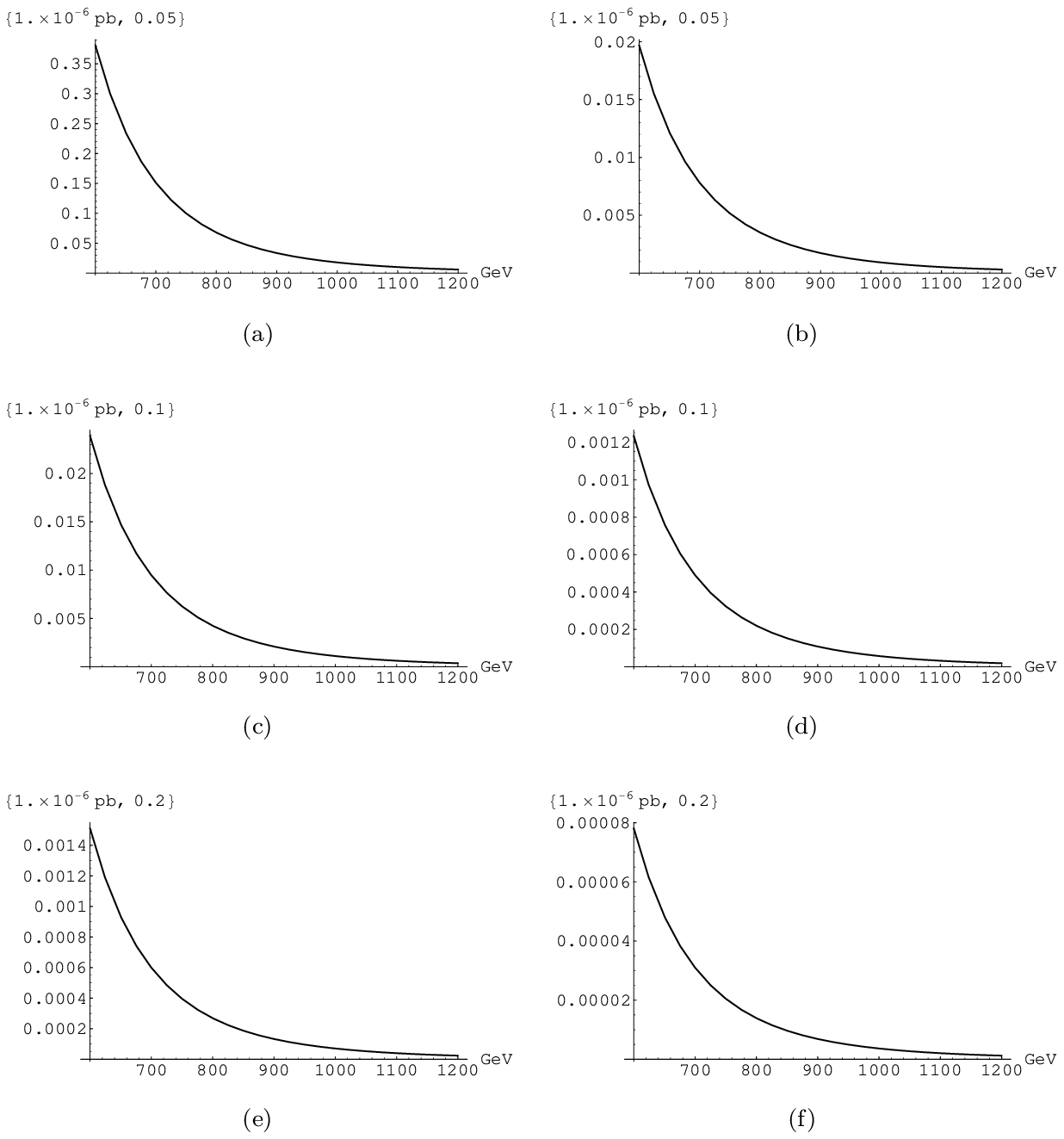}
%\hspace*{0.0cm}\tiny{$(T_A)_{th} \right
%\hspace*{0.0cm}\tiny{$(T_A)_{th} \rightarrow$MeV }
 \caption{The spin proton cross section on the left and that for the neutron on the right in units of
$10^{-6}$pb, as a function of the gauge boson mass in the range of
$600-1200$ GeV. From top to bottom $\Delta =0.05,0.10$ and $0.20$
respectively.} \label{fig2d:spin}
\end{center}
\end{figure}

 Unlike the neutralino case, where one has to live with an allowed SUSY parameter space involving 4 parameters,
see  e.g. Ellis {\it et al} \cite{EOSS04}, Bottino {\it et al},
 and Arnowitt {\it et al} \cite{ref2},
 the situation here is much simpler. One encounters only three mass parameters, $\Delta,m_{B^{(1)}},m_h$,
in the case of the coherent process and only the first two parameters in the case of the spin cross section.
Admittedly, however, the cross section depends on large powers of these masses and, therefore, the
 predictions of the event rates are not very accurate.
The isoscalar spin cross section and the Higgs contribution of the coherent process depend on the structure
of the nucleon. The uncertainties encountered here are no worse than those in the neutralino cross section.
In the evaluation of the parameters $f_q$ one encounters both theoretical and experimental errors. Thus the nucleon cross
section associated with the Higgs mechanism can vary within an order of magnitude \cite{JDV06}. In the present calculation we
will adopt an optimistic approach and employ:
$$f_d=0.041,~f_u=0.028,~f_s=0.400,~f_c=0.051,~f_b=0.055,~f_t=0.095.$$
The thus obtained results for the coherent process are shown in Fig. \ref{fig3d:coh}. The independent variable in
our plots is the mass of the dark matter candidate, since this has become standard in analyzing the experimental searches.
  The cross sections associated with intermediate Higgs scalars only are plotted in Fig. \ref{fig:etah} below.
 In the case of the spin cross section we obtain the 3-dimensional plots shown in Fig. \ref{fig3d:spin}.
Since the cross sections, especially the spin cross sections, are sensitive functions of their arguments we will present the
above results a a series of one dimensional plots. This  is done in Fig. \ref{fig2d:coh} for the coherent mode and
 in Fig. \ref{fig2d:spin} for that of the spin.
%%%%%%%%%%%%%%%%%%%%%%%%%%%%%%%%%%%%%%%%%%%%%%%%%%%%%%%%%%%%%%%%%%
\subsection{The K-K neutrino case}
%%%%%%%%%%%%%%%%%%%%%%%%%%%%%%%%%%%%%%%%%%%%%%%%%%%%%%%%%%%%%%%%%
In this case the expression for the cross section is quite simple.
We will consider each case separately.
%\begin{enumerate}
\subsubsection{ Intermediate $Z$ boson.}
 In this case we have two possibilities:
\begin{itemize}
\item Majorana neutrino.
 Now only the axial current
contributes. The proton and the neutron cross sections are equal
and given by: \beq \sigma_N(spin)=\frac{1}{\pi}\frac{G_F^2}{8}
m^2_p 3 g_A^2=8.0\times 10^{-3}pb .\eeq

\item Dirac neutrino.
In this case  we have again a contribution due to the axial
current, but the resulting nucleon cross section is twice as large
compared to the previous case, i.e.:
    \beq
   \sigma_N(spin)=\frac{1}{\pi}\frac{G_F^2}{8} m^2_p ~3~ 2~ g_A^2=1.6\times 10^{-2}pb
  .
\eeq
 In the case of the neutron we have, however,  an additional
spin independent contribution given by:
    \beq
  \sigma_n(coh)=\frac{1}{\pi}\frac{G_F^2}{8} m^2_p ~ 2~=3.5\times 10^{-3}pb
  .\eeq
  \end{itemize}
  It is quite straightforward to compute the nuclear cross sections:
  \beq
  \sigma_{nuclear}(spin)=\frac{\mu ^2_r }{m^2_p} \sigma_N(spin) \zeta _{spin} F_{11}(q)
  \eeq
  Here $F_{11}$ is the spin response function \cite{DIVA00,JDVSPIN04,JDV06}, which depends on the
energy $Q$ transfered to the nucleus,
 $u=m_AQb^2$, with  $b$ the nuclear harmonic oscillator size parameter,  and $\zeta _{spin}$ is the
nuclear static spin ME given by:
  \beq
 \zeta_{spin}= \frac{1}{3} \left [\Omega_p -\Omega_n \sqrt{\frac{\sigma_n}{\sigma_p}} \right]^2=\frac{1}{3} \left [\Omega_p -\Omega_n  \right]^2
  ,\eeq
 ($\sigma_p=\sigma_n=\sigma_N)$. Here  $\Omega_p$ and $\Omega_n$ are the nuclear spin ME associated with the proton and neutron component respectively.\\
%  and $F_{11}(q)$ is the spin response function . \\
  The coherent cross section becomes
   \beq
  \sigma_{nuclear} (coh)=\frac{\mu ^2_r }{m^2_p} \sigma_n(coh) N^2  \left [ F(q) \right ]^2
  ,\eeq
  where $N$ is the neutron number and $F(q)$ the nuclear form factor.
%  \beq
%  \sigma=\frac{1}{\pi}\frac{G_F^2}{8} \mu^2_r \frac{1}{2 (2J_i+1)}\sum_{m_s's}
%(J^{\lambda}(\nu^{(1)}) J_{\lambda}(NNZ))(J^{\mu}(\nu^{(1)}) J_{\mu}(NNZ))^{*}
%  \eeq
\subsubsection{ The right handed interaction.}
\begin{figure}[!ht]
 \begin{center}
 \includegraphics[scale=1.0]{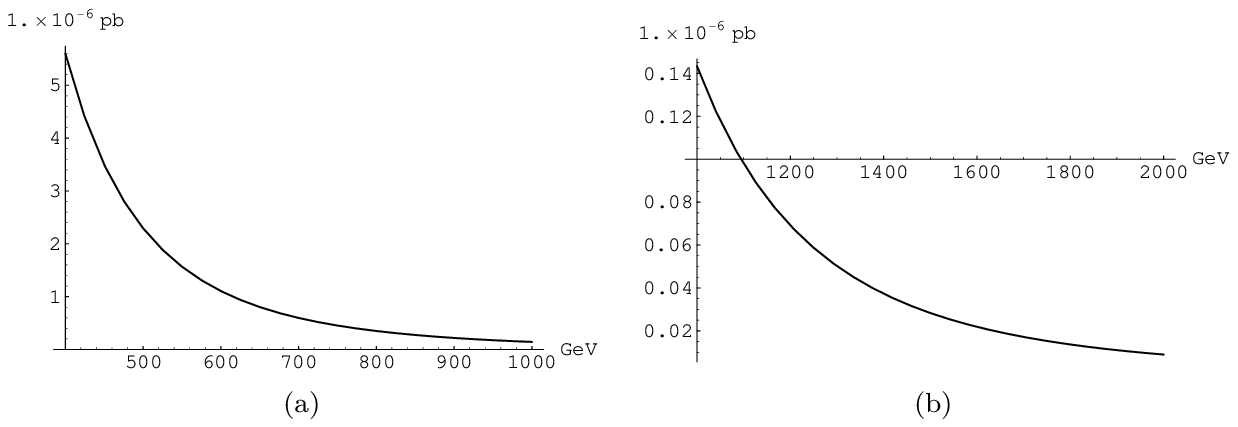}
%\hspace*{0.0cm}\tiny{$(T_A)_{th} \right
%\hspace*{0.0cm}\tiny{$(T_A)_{th} \rightarrow$MeV }
\caption{The coherent nucleon cross section in the case of right
handed neutrino interaction for a Dirac $\nu^{(1)}$ as a function
of the gauge boson mass responsible for this interaction.}
 \label{fig:nur}
  \end{center}
  \end{figure}

In this case case the nucleon cross section is retarded compared to the previous case. The obtained results
are shown in Fig. \ref{fig:nur}.
  
\subsubsection{ The Intermediate Higgs scalar.}

\begin{figure}[!ht]
 \begin{center}
\includegraphics[scale=1.0]{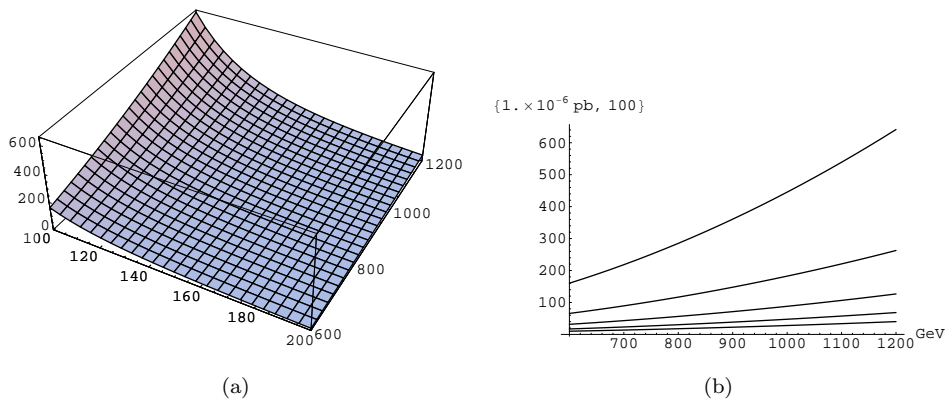}
%\hspace*{0.0cm}\tiny{$(T_A)_{th} \right
%\hspace*{0.0cm}\tiny{$(T_A)_{th} \rightarrow$MeV }
 \caption{On the left we show
the coherent nucleon cross section as a function of $m_{\nu^{(1)}}$ and $m_h$ in GeV.
On the right we show the same thing as a function of the mass of $\nu^{(1)}$ for the indicated Higgs mass
(from top to bottom $100,125,150,175$ and $200$ GeV).
We see that this mechanism excludes a heavy neutrino as a WIMP candidate, unless the Higgs mass is quite large.}
 \label{fig:nuh}
 \end{center}
  \end{figure}

 Naively one expects  this
process to be suppressed due to the small  mass of the u and d
quarks \cite{AGSER05}, present in the nucleon. This is true in the
naive quark model for the nucleon. We have seen above and we know
from the neutralino case that the heavy quarks contribute and in
fact dominate. One finds:
   \beq
  \sigma_N(coh)=\frac{8}{\pi}\left (G_F m^2_p \right ) ^2 \frac{m_p^2 (m_{\nu^{(1)}})^2}{m^4_h}m_p^{-2}(\sum_q f_q)^2 ~=
1.1\times 10^{-1}pb
  \frac{m_p^2 (m_{\nu^{(1)}})^2}{m^4_h}(\sum_q f_q )^2
  .\eeq
The value $\sum_q f_q=0.67$ is acceptable. Using this value we
obtain the results shown in Fig.~\ref{fig:nuh}. We see that this
mechanism excludes a heavy neutrino as a WIMP candidate, unless
the Higgs mass is much larger. In the Standard Model this is
possible and $m_h$ can be treated as a parameter to be extracted
from the data. In SUSY models, however, the lightest neutrino is
expected to be quite light, $m_h\simeq 120$ GeV.
%\end{enumerate}
%%%%%%%%%%%%%%%%%%%%%%%%%%%%%%%%%%%%%%%%%%%%%%%%%%%%
\section{Other non SUSY Models}
\begin{figure}[!ht]
 \begin{center}
 \includegraphics[scale=1.0]{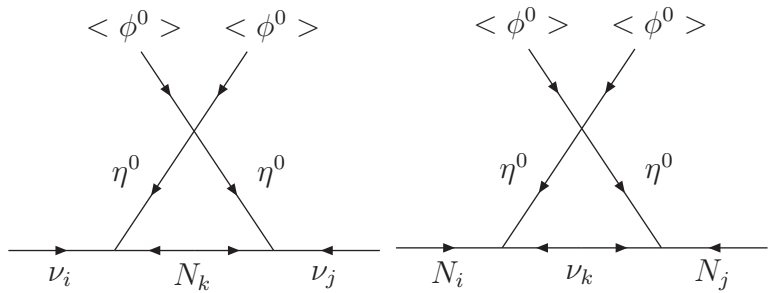}
%  \includegraphics[scale=1.0]{looplightnu.eps}
% \includegraphics[scale=1.0]{loopheavynu.eps}
%\includegraphics[scale=1.0]{fig11.eps}
%\hspace*{0.0cm}\tiny{$(T_A)_{th} \rightarrow$MeV }
 \caption{The 1-loop diagram responsible for the Majorana neutrino masses. Note that, since the isosinglet
neutrino has negative $Z_2$ parity, there is no Dirac mass. An adjustable quartic coupling
$\lambda$ is understood.}
 \label{fig:dirloop}
 \end{center}
  \end{figure}

%%%%%%%%%%%%%%%%%%%%%%%%%%%%%%%%%%%%%%%%%%%%%%%%%%%%%%%
\label{Zmodel}
 There exist extensions of the Standard Model not motivated by symmetry, which may have a particle
content similar to that of K-K theories discussed above. Such models, however, have a much
lower predictive power than the K-K scenarios discussed above. To see this we examine the 
following cases:
\begin{itemize}
\item Models which introduce extra Higgs particles and impose a
discrete symmetry which leads to a "parity" {\it a la} R-parity or
K-K parity \cite{MA06}. \item Extensions of the Standard Model,
which do not require the ad hoc introduction of a parity, but introduce high weak isospin
multiplets \cite{CFA06} with Y=0. So the WIMP-nucleus interaction
via Z-exchange at tree level is absent and the dominant
contribution to the WIMP-nucleus scattering occurs at the one loop
level. 
\end{itemize}
We will consider here the first of the above possibilities
\cite{MA06}. All particles of the Standard Model have parity $+1$,
while the new exotic particles have parity $-1$. The Standard
Model particles of interest to us here, leptons and Higgs, are:
\begin{eqnarray}
&& (\nu_i,l_i) \sim (2,-1/2), ~~~ l^c_i \sim (1,1), ~~~ N_i \sim
(1,0), \\ && (\phi^+,\phi^0) \sim (2,1/2), ~~~ (\eta^+,\eta^0) \sim
(2,1/2).
\end{eqnarray}
in the usual notation $SU(2)_L  \times U(1)_Y $ quantum numbers.
Consider now the following minimal extension of the SM with symmetry $SU(2)_L
\times U(1)_Y \times Z_2$ and  particle content:
\begin{eqnarray}
&& (\nu_i,l_i) \sim (2,-1/2;+), ~~~ l^c_i \sim (1,1;+), ~~~ N_i \sim
(1,0;-), \\ && (\phi^+,\phi^0) \sim (2,1/2;+), ~~~ (\eta^+,\eta^0) \sim
(2,1/2;-).
\end{eqnarray}
Note that the particles $N_i$ and the scalar doublet
$(\eta^+,\eta^0)$ are odd under $Z_2$. This makes the lightest exotic
particle a viable dark matter candidate. In other words in this economic scenario one introduces:
\begin{itemize}
\item  A new doublet of Higgs scalars $\eta$, which have the same  quantum numbers with the ordinary Higgs,
but parity $-1$. These are expected to be quite massive, but they do not develop a vacuum expectation value.
\item Assign parity $-1$ to the usual isosinglet right handed neutrinos.
\end{itemize}
 In this scenario the see-saw mechanism for neutrino mass generation is not operative. One can not have a Dirac mass term
 $\bar{\nu}_{iL} \phi^0 N_{jR}\rightarrow \bar{\nu}_{iL} \prec \phi ^0\succ  N_{jR}$
  since the "parity" forbids it. One can have Majorana mass terms at the one loop level as shown in Fig. \ref{fig:dirloop},
 involving two $\eta$ scalars. The two $\eta$ scalars couple with the ordinary Higgs scalars
with a
 quartic coupling $\lambda$. The net result is that the isosinglet neutrino can be much lighter than that of the
the standard see-saw mechanism.
We now have two possibilities
\begin{itemize}
\item The lightest of the heavy neutrinos is the dark matter candidate.\\
In this case the obtained results are the same as those discussed
in the previous sections in connection with the right handed interaction
(see Fig. \ref{fig:nur}).
 \item The neutral component of the exotic
Higgs scalars $\eta$ is the dark matter candidate. In this case,
since we have not introduced exotic quarks, the interaction of the
dark matter candidate with the quarks is achieved only via the
ordinary Higgs (see Fig.~\ref{fig:etaetah}.
    \begin{figure}[!ht]
 \begin{center}
 \includegraphics[scale=1.0]{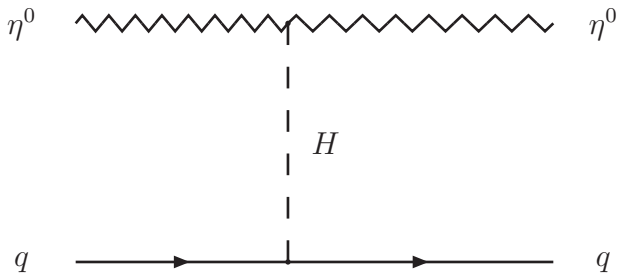}
% \includegraphics[scale=1.0]{etaetah.eps}
%\includegraphics[scale=1.0]{fig12.eps}
%\hspace*{0.0cm}\tiny{$(T_A)_{th} \rightarrow$MeV }
 \caption{The The mechanism leading do direct dark matter detection for a scalar WIMP $\eta^0$, which is stable to
the fact that it has $Z_2$ parity $-1$. This is similar with Fig \ref{fig:kkh}, except that an effective coupling $\lambda_{eff}=\tilde{\lambda}(g^2_1/4)$, with
$\tilde{\lambda}$ treated phenomenologically, is understood.}
 \label{fig:etaetah}
 \end{center}
  \end{figure}
The corresponding effective $\eta \eta h$ coupling is now parameterized as $\lambda_{eff}=\tilde{\lambda}(g^2_1/4)$ ($\tilde{\lambda}=1$
corresponds to the K-K case). Applying the
formalism of the previous section we obtained results like those
 shown in Fig. \ref{fig:etah}.
      \begin{figure}[!ht]
 \begin{center}
 \includegraphics[scale=0.7]{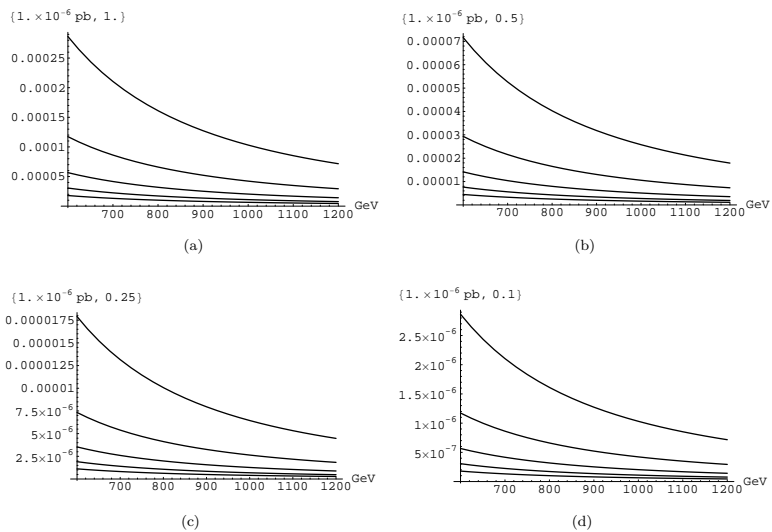}

%\hspace*{0.0cm}\tiny{$(T_A)_{th} \right
 \label{fig:etah} 
%\includegraphics[scale=0.7]{fig13.eps}
%\hspace*{0.0cm}\tiny{$(T_A)_{th} \right
%\hspace*{0.0cm}\tiny{$(T_A)_{th} \rightarrow$MeV }
 \caption{The nucleon cross section corresponding to the $\eta$ scalar being the dark matter candidate,
as a function of the $\eta$ mass. On each graph we present results for $m_{h}=100,125,150,175$ and $200$ GeV (the mass increases downwards).
>From top to bottom and left to right $\tilde{\lambda}=1,~0.5~,~~0.25$ and $0.1$ respectively. Note that
the special case $\tilde{\lambda}=1$ it coincides with the Higgs contribution discussed above in connection with
the K-K theories.}
% \label{fig:etah}
 \end{center}
  \end{figure}
\end{itemize}
As in the case of K-K theories, only in the next generation of experiments such WIMP's can be detected.\\
Before concluding this section we should mention  another interesting extension of the Standard Model
in the direction of technicolour \cite{GKS06}.
In this case the WIMP is %the lightest neutral technibaryon (LTB).
the neutral LTP (lightest neutral technibaryon).
 This is scalar particle, which couples to the quarks via derivative coupling
through Z-exchange. This model, however, in its present form, leads to too
large nucleon cross sections and is excluded by the data.
% \end{itemize}

 \section{Event Rates}
%%%%%%%%%%%%%%%%%%%%%%%%%%%%%%%%%%%%%%%%%%%%%%%%%
 The event rate for the coherent WIMP-nucleus elastic scattering is given by \cite{JDV03,JDV06}:
\beq
R= \frac{\rho (0)}{m_{\chi^0}} \frac{m}{m_p}~
              \sqrt{\langle v^2 \rangle } \left [f_{coh}\left(A,\mu_r(A) \right) \sigma_{p,\chi^0}^{S}+f_{spin}
              \left(A,\mu_r(A)\right)\sigma _{p,\chi^0}^{spin}~\zeta_{spin} \right]
\label{fullrate} ,\eeq with \beq f_{coh}\left(A,Z, \mu_r(A)
\right)=\frac{100\mbox{GeV}}{m_{\chi^0}}\left[
\frac{\mu_r(A)}{\mu_r(p)} \right]^2
\frac{g^2_{coh}(A,Z)}{A}~t_{coh}\left(1+h_{coh}cos\alpha \right)
,\eeq and $g_{coh}(A,Z)=A,A-Z,Z$ for total, neutron and proton
coherence respectively. \beq f_{spin}\left(A,
\mu_r(A)\right)=\frac{100\mbox{GeV}}{m_{\chi^0}} \left[
\frac{\mu_r(A)}{\mu_r(p)} \right]^2
\frac{t_{spin}}{A}\left(1+h_{spin}cos\alpha \right), \eeq
%\frac{\rho(0)}{ \mbox{0.3GeVcm^{-3}}}
with $\sigma_{p,\chi^0}^{S}$ and $\sigma _{p,\chi^0}^{spin}$ the scalar and spin proton cross sections
$~\zeta_{spin}$ the nuclear spin ME.

In this work we will ignore the motion of the earth, i.e. $h_{coh}=h_{spin}=0$ no modulation.
%\begin{enumerate}
\subsection{ The coherent contribution due to the scalar interaction.}

\begin{figure}
\begin{center}
 \includegraphics[scale=1.0]{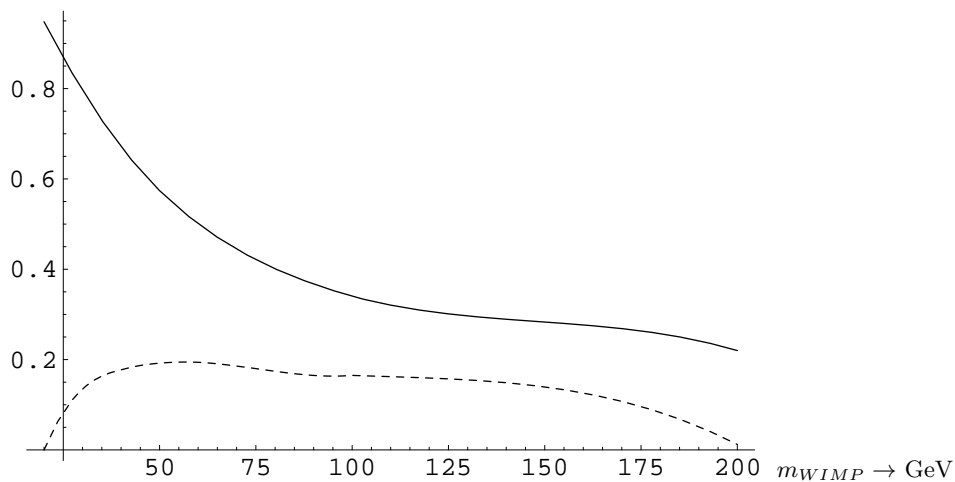}
 \hspace{0.0cm} $m_{WIMP}\rightarrow$ GeV
 \caption{The parameter $t$ in the case of $^{127}$I as a function of the WIMP mass in GeV for zero threshold
(continuous curve) and a threshold of 10 keV (dotted curve). For higher WIMP masses $t$ remains approximately
constant.
}
\label{fig:tI1}
 \end{center}
 \end{figure}

 The number of events in time $t$ due to the scalar interaction, which leads to coherence, is:
\beq
 R\simeq 1.60~10^{-3}
\frac{t}{1 \mbox{y}} \frac{\rho(0)}{ {\mbox0.3GeVcm^{-3}}}
\frac{m}{\mbox{1Kg}}\frac{ \sqrt{\langle v^2 \rangle }}{280 {\mbox
kms^{-1}}}\frac{\sigma_{p,\chi^0}^{S}}{10^{-6} \mbox{ pb}}
f_{coh}(A, Z,\mu_r(A)) \label{eventrate} .\eeq
The parameter $t$
depends on the structure of the nucleus, the WIMP velocity
distribution, the WIMP mass and the energy cutoff imposed by the
detector. In the case of $^{127}$I this parameter is shown in Fig.
\ref{fig:tI1}.
 The nucleon cross section depends on the particle model. We will consider the following cases:
\begin{itemize}
\item The WIMP is a K-K boson \\
In this case we will consider the viable possibility
 $\Delta=0.8$ (see Fig. \ref{fig2d:coh}). Then one obtains the event rates shown in Fig.
 \ref{fig:rI1}.
   \begin{figure}
\begin{center}
 \includegraphics[scale=0.7]{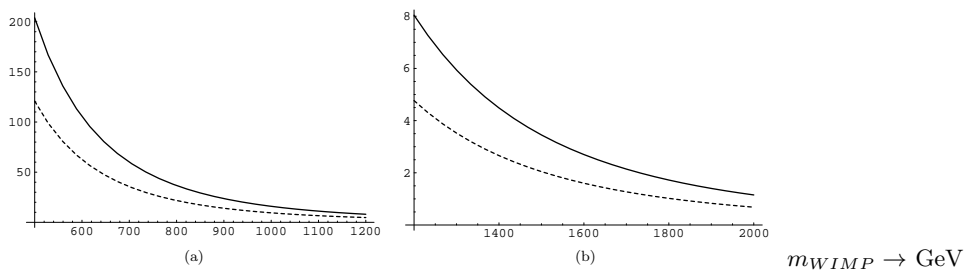}
\hspace{0.0cm} $m_{WIMP}\rightarrow$ GeV \caption{The coherent
event rate $R$ per year per Kg of target in the case of $^{127}$I,
in the case of the K-K gauge boson as WIMP, plotted a as function
of the WIMP mass in GeV for zero threshold (continuous curve) and
a threshold of 10 keV (dotted curve). Both figures show the same
quantity except the WIMP masses range is different. }
\label{fig:rI1}
 \end{center}
 \end{figure}
 We see that, even further from the degeneracy and quite heavy WIMPs, $m_{\chi}\simeq 1$ TeV, the event rates are detectable.
 \item The WIMP is a K-K Majorana neutrino. \\
  The Dirac K-K neutrino case is excluded, since, then, the Z-induced neutron coherent contribution would be too
large. In the case of Majorana neutrinos one can have  coherence
due to the amplitude obtained via the Higgs exchange. The obtained
results are shown in Fig. \ref{fig:rnuh}. From this figure we see
that the lighter Higgs mass is allowed by the data only for
relatively light WIMPs. The heavier Higgs mass, however, is
allowed for all WIMPs. Clearly such heavy Higgs cannot occur in
SUSY theories, since then, $m_h\leq 120$ GeV.
     \begin{figure}
\begin{center}
\includegraphics[scale=1.0]{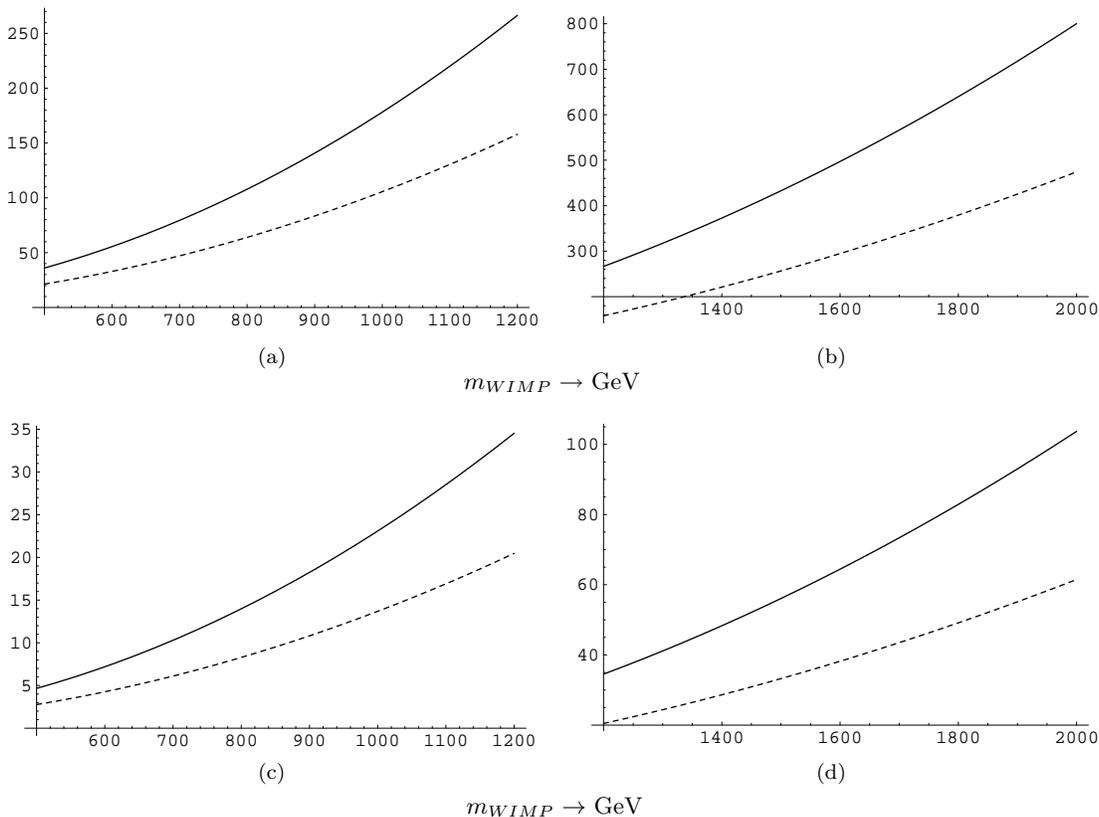}
\hspace{0.0cm} $m_{WIMP}\rightarrow$ GeV
 \caption{The same as in Fig. \ref{fig:rI1} when the WIMP is a K-K Majorana neutrino $\nu^{(1)}$.
 The coherent process is mediated
by Higgs exchange, with $m_h=300$ GeV on the top and 500 GeV at the bottom. The lighter Higgs mass is allowed
by the data only for relatively light WIMPs. The heavier Higgs mass is allowed for all WIMPs. Clearly such heavy Higgs
cannot occur in SUSY theories ($m_h\leq 120$ GeV).
}
\label{fig:rnuh}
 \end{center}
 \end{figure}
 \end{itemize}
\subsection{ The spin interaction.}
In this case, the event in time t rate can be cast into the form:
\beq
 R\simeq 1.60~
\frac{t}{1 \mbox{y}} \frac{\rho(0)}{ {\mbox0.3GeVcm^{-3}}}
\frac{m}{\mbox{1Kg}}\frac{ \sqrt{\langle v^2 \rangle }}{280 {\mbox
kms^{-1}}}\frac{\sigma_{p,\chi^0}^{S}}{10^{-3} \mbox{ pb}}
f_{spin}(A, Z,\mu_r(A)) \zeta_{spin} \label{seventrate} .\eeq Note
that there is a different normalization, since due to the lack of
coherence, the nucleon spin cross section must be larger to yield
detectable results. In the above expression
$$ \zeta_{spin}=\frac{1}{3}\left( \Omega_p+\frac{a_n}{a_p} \Omega_n \right)^2,$$
 with $a_p$, $a_n$ the proton and
 neutron spin amplitudes normalized so that \cite{JDV06} $\sigma_p=|a_p^2|$and
$\sigma_n=|a_n^2|$. $\Omega_p$, $\Omega_n$ are
the nuclear spin matrix elements arising from the protons  and
neutrons respectively, normalized so that $\zeta=1$ for a single
proton. The case of interest to us is when the WIMP is a K-K
Majorana neutrino. In this case we have found that $a_p=-a_n$ and
$\sigma_p=\sigma_n=8.0 \ 10^{-3}$ pb. We will examine the
following cases:
\begin{itemize}
\item The target $^{19}$F ($\Omega_p=1.646,\Omega_n=0.30$).\\
 This light target is favored from the spin ME point of view \cite{DIVA00,JDV06}, but for heavy WIMPs is disfavored due to
the small reduced mass. The parameter $t_{spin}$ is shown in Fig.
\ref{fig:tsF}. The obtained rates are shown in Fig. \ref{fig:rsF}.
  \begin{figure}
\begin{center}
 \includegraphics[scale=1.0]{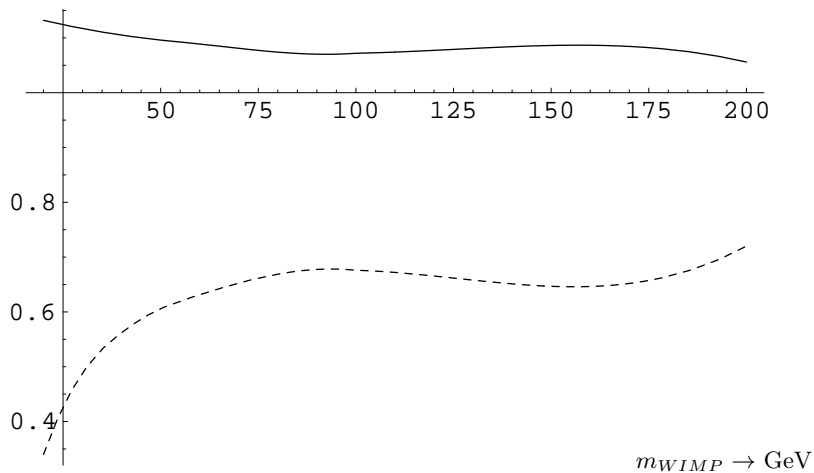}
 \hspace{-2.0cm} $m_{WIMP}\rightarrow$ GeV
 \caption{The parameter $t_{spin}$ in the case of $^{19}$F as a function of the WIMP mass in GeV for zero threshold
(continuous curve) and a threshold of 10 keV (dotted curve). For higher WIMP masses $t_{spin}$ remains approximately
constant.
}
\label{fig:tsF}
 \end{center}
 \end{figure}
    \begin{figure}
\begin{center}
 \includegraphics[scale=1.0]{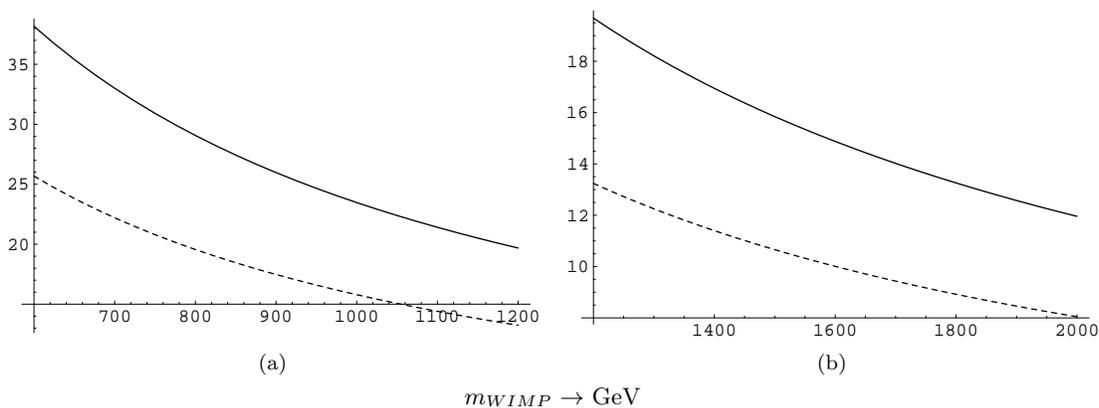}
 \hspace{0.0cm} $m_{WIMP}\rightarrow$ GeV
 \caption{The spin event rate $R$ per year per Kg of target in the case of $^{19}$F, in the cases the WIMP
is a K-K majorana neutrino, plotted  as a function of the
WIMP mass in GeV for zero threshold
(continuous curve) and a threshold of 10 keV (dotted curve). Both figures show the same quantity except the
 WIMP masses range is different.
}
\label{fig:rsF}
 \end{center}
 \end{figure}
 \item The target $^{73}$Ge ($\Omega_p=0.036,\Omega_n=1.040$).\\
 This medium mass  target,  favored for the coherent process as well, is characterized by large
spin \cite{JDV06},\cite{Ress}-\cite{SUHONEN03} induced rates.
The parameter $t_{spin}$ is shown in Fig. \ref{fig:tsGe}. The
obtained rates are shown in Fig. \ref{fig:rsGe}
  \begin{figure}
\begin{center}
 \includegraphics[scale=1.0]{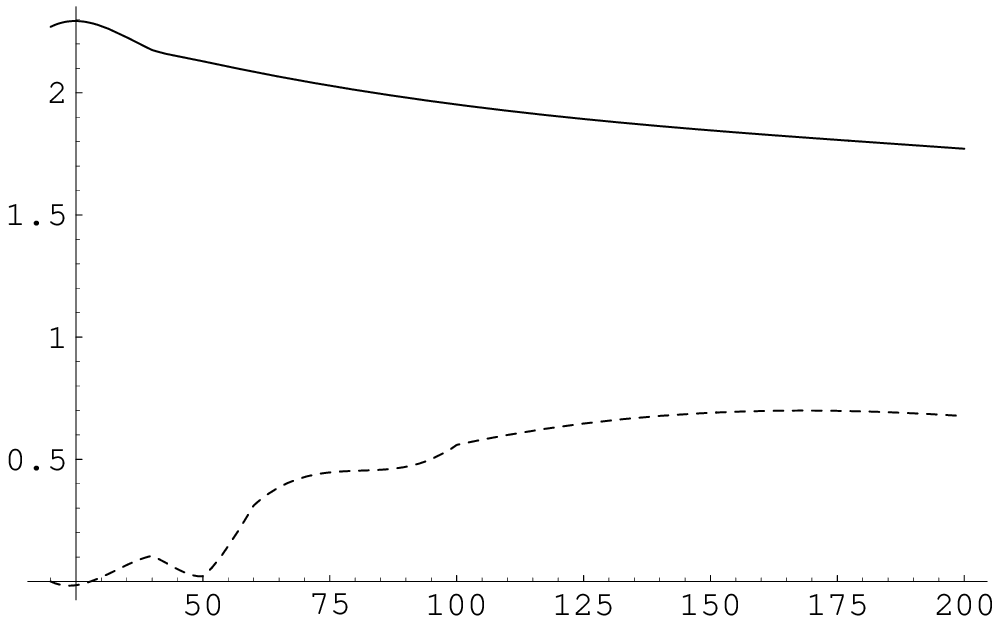}
 \hspace{0.0cm} $m_{WIMP}\rightarrow$ GeV
 \caption{The parameter $t_{spin}$ in the case of $^{73}$Ge. Otherwise the notation is the same as in Fig. \ref{fig:tsF}.
}
\label{fig:tsGe}
 \end{center}
 \end{figure}
    \begin{figure}
\begin{center}
 \includegraphics[scale=1.0]{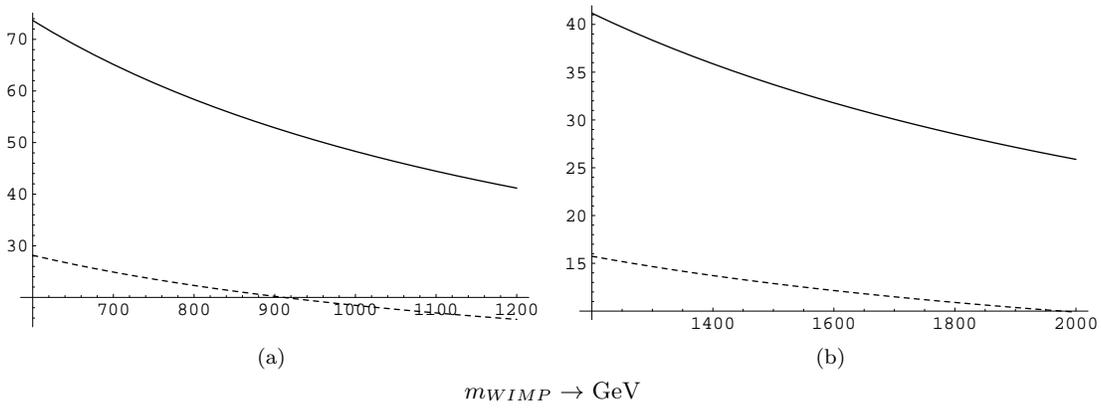}
 \hspace{0.0cm} $m_{WIMP}\rightarrow$ GeV
 \caption{The same as in \ref{fig:rsF} in the case of the $^{73}$Ge target.
}
\label{fig:rsGe}
 \end{center}
 \end{figure}
  \item The target $^{127}$I ($\Omega_p=1.460,\Omega_n=0.355$).\\
 This medium mass  target, with which the DAMA  experiment \cite{BERNA2,BERNA1} claimed to have  observed a signal,
is favored for the spin contribution as well due to the large
reduced mass, even though the spin ME is modest \cite{Ress}-\cite{SUHONEN03}. The parameter
$t_{spin}$ is shown in Fig. \ref{fig:tsI}. The obtained rates are
shown in Fig. \ref{fig:rsI}
  \begin{figure}
\begin{center}
 \includegraphics[scale=1.0]{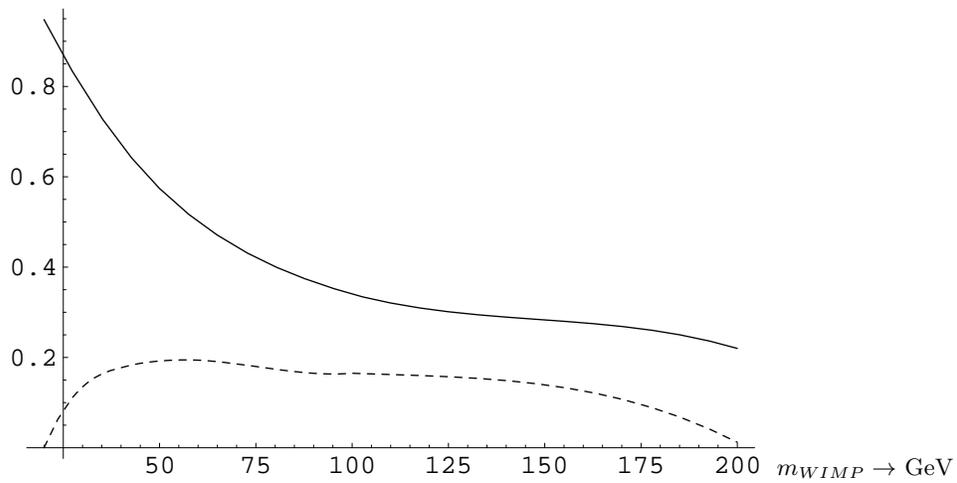}
 \hspace{0.0cm} $m_{WIMP}\rightarrow$ GeV
 \caption{The parameter $t_{spin}$ in the case of $^{127}$I. Otherwise the notation is the same as in Fig. \ref{fig:tsF}.
}
\label{fig:tsI}
 \end{center}
 \end{figure}
    \begin{figure}
\begin{center}
\includegraphics[scale=1.0]{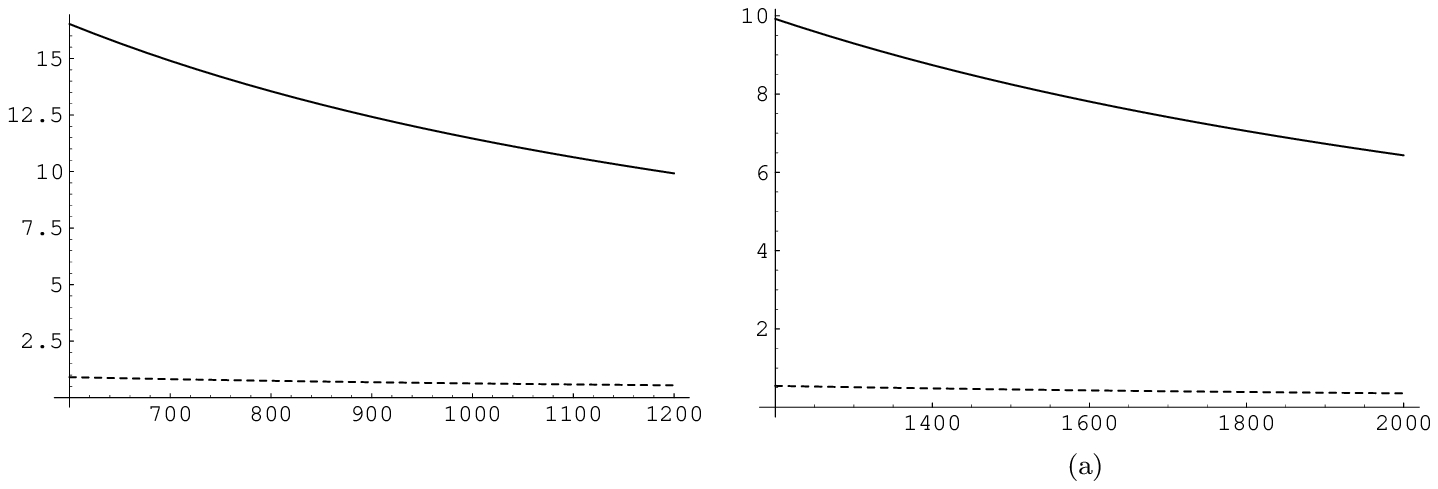}
\hspace{0.0cm} $m_{WIMP}\rightarrow$ GeV
 \caption{The same as in \ref{fig:rsF} in the case of the $^{127}$I target.
}
\label{fig:rsI}
 \end{center}
 \end{figure}
 We see that a K-K Majorana neutrino is not excluded by the data.
%The spin contribution induced via Z-exchange is comparable
%to the coherent, unless the Higgs is light.
 \end{itemize}
%\end{enumerate}
%with

%$\gamma _{lambda} \left[  t_3(q)(1-\gamma  _5)-2 e_q^2 2 \sin^2 {\theta} \right] q$
%%%%%%%%%%%%%%%%%%%%%%%%%%%%%%%%%%%%%%%%%%%%%%%%%%
\section{Discussion}
%%%%%%%%%%%%%%%%%%%%%%%%%%%%%%%%%%%%%%%%%%%%%%%%%%%%
 Even though the neutralino is the preferred WIMP candidate, in this work we concentrated on non-SUSY WIMPs.
% In theories involving scalar WIMPs, like the technicolor theories \cite{GKS06}, for a fiven amplitude,
%even if it is a typical
%week interaction size, the cross section falls as $1/m^2_{\chi}$ and the rate as  $1/m^3_{\chi}$.
%. So one does not expect sizable rates for very heavy WIMPs.
 Extensions of the Standard Model, not motivated by some symmetry, involve many parameters and do not
 have much predictive power. So we will concentrate on K-K WIMPs, whereby the couplings involved are those
of the Standard Model. Thus essentially one encounters only one
unknown parameter,
 namely the WIMP mass.

>From the results of the previous section, in connection with the K-K WIMPs as dark matter candidates,
 one can conclude the following:
\begin{itemize}
\item The K-K neutrinos as CDM candidate.\\
In this case everything is under control, except of course, the
fact that we do not know for sure whether the K-K neutrinos are
Majorana  or Dirac particles. Most authors expect them to be Dirac
neutrinos (see, e.g. Servant\cite{SERVANT}). In the case of Higgs
contribution the nucleon cross section is the same both for Dirac
and Majorana neutrinos. It is proportional to the $ [m_{\nu^{(1)}}
]^2$ and excludes the K-K neutrino as a viable WIMP candidate,
unless the lightest Higgs is very heavy. In all other cases  the
WIMP mass enters  via $\mu_r$, both explicitly and implicitly
through the nuclear form factor. Anyway, since, essentially  from cosmological
requirements \cite{ST02a,ST02b}, the K-K neutrino mass is
expected to be in the TeV region, $\mu_r\simeq A m_p$.
 So the cross section is independent of the WIMP mass. The
number of WIMPs in our vicinity, for a given density, is inversely
proportional to the WIMP mass. Thus the rate will scale as
follows: \beq R(m_{WIMP})=R(A)\frac{A {\mbox GeV }}{m_{WIMP}}
\label{largem} .\eeq
 Large nucleon spin cross sections ($\sim 10^{-2}$) pb  are possible via Z-exchange. Thus Dirac neutrinos are excluded
due to neutron coherence. No such coherence exists for Majorana neutrinos, so these can not  be excluded from the
data  (16 events per year per Kg of target\cite{CDMS04}). The precise predicted rates depend on nuclear physics assumptions.
%The mechanism of the $\nu^{(1)}\rightarrow \nu$ conversion, discussed above, seems to exclude the K-K neutrino as a WIMP candidate.
% Anyway from a detailed
% analysis of the data one can set lower bounds on the K-K neutrino mass.
\item The K-K boson as CDM candidate.\\
For heavy WIMPs Eq. (\ref{largem}) holds in this case as well allowing larger nucleon cross-sections to be consistent with
the data.
\begin{enumerate}
\item The unknown parameters of the theory are the masses of K-K
quarks and gauge bosons as well as the mass of the neutral Higgs.
All relevant couplings are under control. \item In the spin
independent mechanism the proton cross section is dominant. This
is due to the RL+LR currents. This prediction can be consistent
with the present data only away from the resonance and/or large
K-K gauge boson masses. We should also  take note of the fact that
the event rate will be down by a factor $Z^2/A^2$ compared to the
analysis of the neutralino case. This contribution (LR+RL) is
absent in the spin mode. It is also absent in the case of the
neutron cross section. \item Even in the other cases the proton
cross sections are larger than those for the  neutrons. \item The
results fall quite fast with increasing boson mass. \item The
obtained results, in particular those associated with the spin,
are very sensitive functions of the mass difference $\Delta$
between the K-K quarks and the K-K bosons. \item For sufficiently
small $\Delta$, the process involving the K-K quarks is more
important than the Higgs induced cross section. Away from the
resonance the Higgs contribution becomes significant.
%\item Since the spin nucleon cross section is not much larger than that of the
% coherent mechanism, it is not so relevant.
%\item Only for the light values of the gauge boson mass the cross section is close to the limits extracted
%from the present experiments. For higher masses the cross section may be undetectable even by the next
%generation of experiments
\end{enumerate}
\end{itemize}
\section*{Acknowledgments}  One of the authors (JDV) is indebted to
Ignatios Antoniadis and Geraldine Servant for discussions during his visit at CERN. His
work and this visit were supported by European Union under the
contract MRTN-CT-2004-503369 as well as the program PYTHAGORAS-1.
The latter is part of the Operational Program for Education and
Initial Vocational Training of the Hellenic Ministry of Education
under the 3rd Community Support Framework and the European Social
Fund. The other two authors (V.K.  Oikonomou and ChCM) acknowledge support by the
 co-funded European
Union-European Social fund and National fund PYTHAGORAS -EPEAEK II.
%\bibliography{TeX}

\begin{thebibliography}{34}
\expandafter\ifx\csname natexlab\endcsname\relax\def\natexlab#1{#1}\fi
\expandafter\ifx\csname bibnamefont\endcsname\relax
  \def\bibnamefont#1{#1}\fi
\expandafter\ifx\csname bibfnamefont\endcsname\relax
  \def\bibfnamefont#1{#1}\fi
\expandafter\ifx\csname citenamefont\endcsname\relax
  \def\citenamefont#1{#1}\fi
\expandafter\ifx\csname url\endcsname\relax
  \def\url#1{\texttt{#1}}\fi
\expandafter\ifx\csname urlprefix\endcsname\relax\def\urlprefix{URL }\fi
\providecommand{\bibinfo}[2]{#2}
\providecommand{\eprint}[2][]{\url{#2}}

\bibitem[{\citenamefont{G.~Servant}(2003)}]{ST02a}
\bibinfo{author}{\bibfnamefont{T.~M. P.~T.} \bibnamefont{G.~Servant}},
  \bibinfo{journal}{Nuc. Phys. B} \textbf{\bibinfo{volume}{650}},
  \bibinfo{pages}{391} (\bibinfo{year}{2003}).

\bibitem[{\citenamefont{G.~Servant}(2002)}]{ST02b}
\bibinfo{author}{\bibfnamefont{T.~M. P.~T.} \bibnamefont{G.~Servant}},
  \bibinfo{journal}{New Jour. Phys.} \textbf{\bibinfo{volume}{4}},
  \bibinfo{pages}{99} (\bibinfo{year}{2002}).

\bibitem[{\citenamefont{I.~Antoniadis}(1998)}]{Antoniadis-a}
\bibinfo{author}{\bibfnamefont{S.~D.} \bibnamefont{I.~Antoniadis},
  \bibfnamefont{N.~Arkani-Hamed}}, \bibinfo{journal}{Phys. Lett. B}
  \textbf{\bibinfo{volume}{436}}, \bibinfo{pages}{267} (\bibinfo{year}{1998})
  \bibinfo{note}{; arXiv:hep-ph/9804398.}

\bibitem[{\citenamefont{N.~Arkani-Hamed}(1999)}]{Arkani-b}
\bibinfo{author}{\bibfnamefont{G.~R.~D.} \bibnamefont{N.~Arkani-Hamed},
  \bibfnamefont{S.~Dimopoulos}}, \bibinfo{journal}{Phys. Rev. D}
  \textbf{\bibinfo{volume}{59}}, \bibinfo{pages}{086004}
  (\bibinfo{year}{1999}) \bibinfo{note}{; arXiv:hep-ph/9807344.}

\bibitem[{\citenamefont{K.~R.~Dienes}(1999)}]{Dienes-c}
\bibinfo{author}{\bibfnamefont{T.~G.} \bibnamefont{K.~R.~Dienes},
  \bibfnamefont{E.~Dudas}}, \bibinfo{journal}{Nucl. Phys. B}
  \textbf{\bibinfo{volume}{47}}, \bibinfo{pages}{537} (\bibinfo{year}{1999})
  \bibinfo{note}{; arXiv:hep-ph/9806292.}

\bibitem[{\citenamefont{T.~Appelquist}(2001)}]{Appelquist-d}
\bibinfo{author}{\bibfnamefont{B.~A.~D.} \bibnamefont{T.~Appelquist},
  \bibfnamefont{H.~C.~Cheng}}, \bibinfo{journal}{Phys. Rev. D}
  \textbf{\bibinfo{volume}{64}}, \bibinfo{pages}{035002}
  (\bibinfo{year}{2001}) \bibinfo{note}{; arXiv: hep-ph/0012100.}

\bibitem[{\citenamefont{I.~Antoniadis}(1999)}]{Antoniadis-g}
\bibinfo{author}{\bibfnamefont{M.~Q.} \bibnamefont{I.~Antoniadis},
  \bibfnamefont{S.~Dimopoulos}}, \bibinfo{journal}{Nucl. Phys B}
  \textbf{\bibinfo{volume}{544}}, \bibinfo{pages}{503} (\bibinfo{year}{1999})
  \bibinfo{note}{; arXiv: hep-ph/9810410.}

\bibitem[{\citenamefont{I.~Antoniadis}(1994)}]{Antoniadis-h}
\bibinfo{author}{\bibfnamefont{M.~Q.} \bibnamefont{I.~Antoniadis},
  \bibfnamefont{K.~Benakli}}, \bibinfo{journal}{Phys. Lett. B}
  \textbf{\bibinfo{volume}{331}}, \bibinfo{pages}{313} (\bibinfo{year}{1994})
  \bibinfo{note}{; arXiv: hep-ph/9403290.}

\bibitem[{SER()}]{SERVANT}
\bibinfo{note}{See, e.g., G. Servant, in Les Houches :Physics at TeV Colliders
  2005'' Beyond the Standard Model working group: summary report, B.C. Allanach
  (ed.), C. Grojean (ed.), P. Skands (ed.), {\it al}, section 25, p. 164;
  hep-ph/0602198.}

\bibitem[{MAX()}]{MAXIMA-1}
\bibinfo{note}{S. Hanary {\it et al}: {\it Astrophys. J.} {\bf 545}, L5
  (2000);\\ J.H.P Wu {\it et al}: {\it Phys. Rev. Lett.} {\bf 87}, 251303
  (2001);\\ M.G. Santos {\it et al}: {\it Phys. Rev. Lett.} {\bf 88}, 241302
  (2002)}.

\bibitem[{BOO()}]{BOOMERANG}
\bibinfo{note}{P. D. Mauskopf {\it et al}: {\it Astrophys. J.} {\bf 536}, L59
  (2002);\\ S. Mosi {\it et al}: {\it Prog. Nuc.Part. Phys.} {\bf 48}, 243
  (2002);\\ S. B. Ruhl {\it al}, astro-ph/0212229 and references therein.}

\bibitem[{DAS()}]{DASI}
\bibinfo{note}{N. W. Halverson {\it et al}: {Astrophys. J.} {\bf 568}, 38
  (2002)\\ L. S. Sievers {\it et al}: astro-ph/0205287 and references therein.}

\bibitem[{\citenamefont{{Smoot et al (COBE Collaboration)}}(1992)}]{COBE}
\bibinfo{author}{\bibfnamefont{G.~F.} \bibnamefont{{Smoot et al (COBE
  Collaboration)}}}, \bibinfo{journal}{Astrophys. J.}
  \textbf{\bibinfo{volume}{396}}, \bibinfo{pages}{L1} (\bibinfo{year}{1992}).

\bibitem[{\citenamefont{{Spergel et al}}(2003)}]{SPERGEL}
\bibinfo{author}{\bibfnamefont{D.~N.} \bibnamefont{{Spergel et al}}},
  \bibinfo{journal}{Astrophys. J. Suppl.} \textbf{\bibinfo{volume}{148}},
  \bibinfo{pages}{175} (\bibinfo{year}{2003}).

\bibitem[{\citenamefont{{Tegmark et al}}(2004)}]{SDSS}
\bibinfo{author}{\bibfnamefont{M.}~\bibnamefont{{Tegmark et al}}},
  \bibinfo{journal}{Phys.Rev. D} \textbf{\bibinfo{volume}{69}},
  \bibinfo{pages}{103501} (\bibinfo{year}{2004}).

\bibitem[{WMA()}]{WMAP06}
\bibinfo{note}{D.N. Spergel {\it et al}, Three-Year WMAP Results: Implications
  for Cosmology, astro-ph/0603449;\\ L. Page {\it et al}, Three-Year WMAP
  Results: Polarization Analysis, astro-ph/0603450;\\ G. Hinsaw {\it et al},
  Three-Year WMAP Observations: Implications Temperature Analysis,
  astro-ph/0603451;\\ N Jarosik {\it et al}, Three-Year WMAP Observations: Beam
  Profiles, Data Processing, Radiometer Characterization and Systematic Error
  Limits, astro-ph/0603452.}

\bibitem[{Dre()}]{Dree00}
\bibinfo{note}{A. Djouadi and M. K. Drees, {\it Phys. Lett. B} {\bf 484}, 183
  (2000); S. Dawson, {\it Nucl. Phys. B} {\bf 359}, 283 (1991); M. Spira {it et
  al}, {\it Nucl. Phys.} {\bf B453}, 17 (1995).}

\bibitem[{Che()}]{Chen}
\bibinfo{note}{T. P. Cheng, {\it Phys. Rev. D} {\bf 38}, 2869 (1988); H-Y.
  Cheng, {\it Phys. Lett. B} {\bf 219}, 347 (1989).}

\bibitem[{JDV()}]{JDV06}
\bibinfo{note}{See, e.g., our recent review: J.D. Vergados, On the direct
  detection of dark matter- Exploring all the signatures of the
  neutralino-nucleus interaction, hep-ph/0601064 and references therein.}

\bibitem[{JEL()}]{JELLIS}
\bibinfo{note}{The Strange Spin of the Nucleon, J. Ellis and M. Karliner,
  hep-ph/9501280.}

\bibitem[{\citenamefont{Ellis et~al.}(2004)\citenamefont{Ellis, Olive, Santoso,
  and Spanos}}]{EOSS04}
\bibinfo{author}{\bibfnamefont{J.}~\bibnamefont{Ellis}},
  \bibinfo{author}{\bibfnamefont{K.~A.} \bibnamefont{Olive}},
  \bibinfo{author}{\bibfnamefont{Y.}~\bibnamefont{Santoso}}, \bibnamefont{and}
  \bibinfo{author}{\bibfnamefont{V.~C.} \bibnamefont{Spanos}},
  \bibinfo{journal}{Phys.Rev. D} \textbf{\bibinfo{volume}{70}},
  \bibinfo{pages}{055005} (\bibinfo{year}{2004}).

\bibitem[{ref()}]{ref2}
\bibinfo{note}{A. Bottino {\it et al.}, {\it Phys. Lett B} {\bf 402}, 113
  (1997).\\ R. Arnowitt. and P. Nath, {\it Phys. Rev. Lett.} {\bf 74}, 4592
  (1995); {\it Phys. Rev. D} {\bf 54}, 2374 (1996); hep-ph/9902237;\\ V. A.
  Bednyakov, H.V. Klapdor-Kleingrothaus and S.G. Kovalenko, {\it Phys. Lett. B}
  {\bf 329}, 5 (1994).}

\bibitem[{\citenamefont{Divari et~al.}(2000)\citenamefont{Divari, Kosmas,
  Vergados, and Skouras}}]{DIVA00}
\bibinfo{author}{\bibfnamefont{P.~C.} \bibnamefont{Divari}},
  \bibinfo{author}{\bibfnamefont{T.~S.} \bibnamefont{Kosmas}},
  \bibinfo{author}{\bibfnamefont{J.~D.} \bibnamefont{Vergados}},
  \bibnamefont{and} \bibinfo{author}{\bibfnamefont{L.~D.}
  \bibnamefont{Skouras}}, \bibinfo{journal}{Phys. Rev. C}
  \textbf{\bibinfo{volume}{61}}, \bibinfo{pages}{054612}
  (\bibinfo{year}{2000}).

\bibitem[{\citenamefont{Vergados}(2004)}]{JDVSPIN04}
\bibinfo{author}{\bibfnamefont{J.~D.} \bibnamefont{Vergados}},
  \bibinfo{journal}{J.Phys. G} \textbf{\bibinfo{volume}{30}},
  \bibinfo{pages}{1127} (\bibinfo{year}{2004}), \bibinfo{note}{0406134}.

\bibitem[{\citenamefont{G.~Agashe}(2005)}]{AGSER05}
\bibinfo{author}{\bibfnamefont{G.~S.} \bibnamefont{G.~Agashe}},
  \bibinfo{journal}{JCAP} \textbf{\bibinfo{volume}{0502}}, \bibinfo{pages}{002}
  (\bibinfo{year}{2005}) \bibinfo{note}{; arXiv:hep-ph/0411254.}

\bibitem[{\citenamefont{Ma}(2006)}]{MA06}
\bibinfo{author}{\bibfnamefont{E.}~\bibnamefont{Ma}},
  \bibinfo{journal}{Mod.Phys.Lett. A} \textbf{\bibinfo{volume}{21}},
  \bibinfo{pages}{1777} (\bibinfo{year}{2006}) \bibinfo{note}{;
  hep-ph/0605180.}

\bibitem[{CFA()}]{CFA06}
\bibinfo{note}{M. Cirelli, N. Forengo and A. Sturmia, Minimal Dark Matter,
  hep-ph/0512090 (to appear in Nuc. Phys.B 753).}

\bibitem[{\citenamefont{F.~Sannino}(2005)}]{GKS06}
\bibinfo{author}{\bibfnamefont{K.~T.} \bibnamefont{F.~Sannino}},
  \bibinfo{journal}{Phys. Rev. D} \textbf{\bibinfo{volume}{71}},
  \bibinfo{pages}{2005} (\bibinfo{year}{2005}) \bibinfo{note}{; S.V.
  Gudnason,C. Kouvaris and F Sannino, Dark Matter from New Technicolour
  Theories, hep-ph/0608055;}.

\bibitem[{\citenamefont{Vergados}(2003)}]{JDV03}
\bibinfo{author}{\bibfnamefont{J.~D.} \bibnamefont{Vergados}},
  \bibinfo{journal}{Phys. Rev. D} \textbf{\bibinfo{volume}{57}},
  \bibinfo{pages}{103003} (\bibinfo{year}{2003})
  \bibinfo{note}{;hep-ph/0303231}.

\bibitem[{Res()}]{Ress}
\bibinfo{note}{M. T. Ressell {\it et al.}, {\it Phys. Rev. D} {\bf 48}, 5519
  (1993); M.T. Ressell and D. J. Dean, Phys. Rev. C {\bf 56}, 535 (1997).}

\bibitem[{SUH()}]{SUHONEN03}
\bibinfo{note}{E. Homlund and M. Kortelainen and T. S. Kosmas and J. Suhonen
  and J. Toivanen, Phys. Lett B, {\bf 584},31 (2004); Phys. Atom. Nucl. {\bf
  67}, 1198 (2004).}

\bibitem[{\citenamefont{{Bernabei et al}}(1996)}]{BERNA2}
\bibinfo{author}{\bibfnamefont{R.}~\bibnamefont{{Bernabei et al}}},
  \bibinfo{journal}{Phys. Lett. B} \textbf{\bibinfo{volume}{389}},
  \bibinfo{pages}{757} (\bibinfo{year}{1996}).

\bibitem[{\citenamefont{{Bernabei et al}}(1998)}]{BERNA1}
\bibinfo{author}{\bibfnamefont{R.}~\bibnamefont{{Bernabei et al}}},
  \bibinfo{journal}{Phys. Lett. B} \textbf{\bibinfo{volume}{424}},
  \bibinfo{pages}{195} (\bibinfo{year}{1998}).

\bibitem[{\citenamefont{et~al (CDMS~collaboration)}(2004)}]{CDMS04}
\bibinfo{author}{\bibfnamefont{D.~A.} \bibnamefont{et~al
  (CDMS~collaboration)}}, \bibinfo{journal}{Phys. Rev. Let.}
  \textbf{\bibinfo{volume}{93}}, \bibinfo{pages}{211301}
  (\bibinfo{year}{2004}).

\end{thebibliography}

\end{document}